\pdfoutput=1
\documentclass[12pt,a4paper]{article}
\usepackage{ifthen} % for conditional statements
\newboolean{pdflatex}
\setboolean{pdflatex}{true} % False for eps figures 

\newboolean{articletitles}
\setboolean{articletitles}{true} % False removes titles in references

\newboolean{uprightparticles}
\setboolean{uprightparticles}{false} %True for upright particle symbols

\def\paperauthors{LHCb collaboration} % Leave as is for PAPER, CONF and FIGURE
\def\paperasciititle{Observation of the charmless purely baryonic decay Lb to L p pbar} % Set ASCII title here !! MAKE sure it's only ASCII characters !! 
\def\papertitle{Observation of the charmless\\purely baryonic decay $\mathinner{\mathit{\Lambda}^0_b\!\to \mathit{\Lambda} p \overline{p}}$} % Latex formatted title
\def\paperkeywords{{High Energy Physics}, {LHCb}} % Comma separated list
\def\papercopyright{\the\year\ CERN for the benefit of the LHCb collaboration} % new since 9/Apr/2018
\def\paperlicence{CC BY 4.0 licence}
\def\paperlicenceurl{https://creativecommons.org/licenses/by/4.0/}

\newif\ifEnableSectionTOCLinks
\EnableSectionTOCLinksfalse % deactivated

\usepackage[top=1in, bottom=1.25in, left=1in, right=1in]{geometry}

\usepackage{microtype}
\usepackage{lineno}  % for line numbering during review
\usepackage{xspace} % To avoid problems with missing or double spaces after
\usepackage{caption} %these three command get the figure and table captions automatically small

\usepackage{subcaption}

\usepackage{graphicx}  % to include figures (can also use other packages)
\usepackage{color}
\usepackage{colortbl}
\graphicspath{{./figs/}} % Make Latex search fig subdir for figures
\usepackage{amsmath} % Adds a large collection of math symbols
\usepackage{amssymb}
\usepackage{amsfonts}
\usepackage{upgreek} % Adds in support for greek letters in roman typeset

\newcommand*\patchAmsMathEnvironmentForLineno[1]{%
\expandafter\let\csname old#1\expandafter\endcsname\csname #1\endcsname
\expandafter\let\csname oldend#1\expandafter\endcsname\csname
end#1\endcsname
 \renewenvironment{#1}%
   {\linenomath\csname old#1\endcsname}%
   {\csname oldend#1\endcsname\endlinenomath}%
}
\newcommand*\patchBothAmsMathEnvironmentsForLineno[1]{%
  \patchAmsMathEnvironmentForLineno{#1}%
  \patchAmsMathEnvironmentForLineno{#1*}%
}
\AtBeginDocument{%
\patchBothAmsMathEnvironmentsForLineno{equation}%
\patchBothAmsMathEnvironmentsForLineno{align}%
\patchBothAmsMathEnvironmentsForLineno{flalign}%
\patchBothAmsMathEnvironmentsForLineno{alignat}%
\patchBothAmsMathEnvironmentsForLineno{gather}%
\patchBothAmsMathEnvironmentsForLineno{multline}%
\patchBothAmsMathEnvironmentsForLineno{eqnarray}%
}

\usepackage[pdftex,
            pdfauthor={\paperauthors},
            pdftitle={\paperasciititle},
            pdfkeywords={\paperkeywords}]{hyperref}
\usepackage{hyperxmp}
\hypersetup{
    pdfcopyright={Copyright (C) \papercopyright},
    pdflicenseurl={\paperlicenceurl}
}
\usepackage[colorinlistoftodos,textsize=scriptsize]{todonotes}

\usepackage[bottom,flushmargin,hang,multiple]{footmisc}

\usepackage[all]{hypcap} % Internal hyperlinks to floats.

\usepackage{xspace} 
\usepackage{upgreek}

\def\lhcb   {\mbox{LHCb}\xspace}

\def\MagUp {\mbox{\em Mag\kern -0.05em Up}\xspace}

\ifthenelse{\boolean{uprightparticles}}%
{

 \def\Ppi         {\ensuremath{\uppi}\xspace}

 \def\Pchi        {\ensuremath{\upchi}\xspace}

 \def\PDelta      {\ensuremath{\Delta}\xspace}                 
 \def\PXi         {\ensuremath{\Xi}\xspace}                 
 \def\PLambda     {\ensuremath{\Lambda}\xspace}                 
 \def\PSigma      {\ensuremath{\Sigma}\xspace}                 
 \def\POmega      {\ensuremath{\Omega}\xspace}                 
 \def\PUpsilon    {\ensuremath{\Upsilon}\xspace}
 \let\oldPi\Pi
 \def\PPi         {\ensuremath{\oldPi}\xspace}

 \def\PB      {\ensuremath{\mathrm{B}}\xspace}                 
 \def\PD      {\ensuremath{\mathrm{D}}\xspace}                 

 \def\PK      {\ensuremath{\mathrm{K}}\xspace}                 
 \def\Pb      {\ensuremath{\mathrm{b}}\xspace}                 
 \def\Pc      {\ensuremath{\mathrm{c}}\xspace}

 \def\Pp      {\ensuremath{\mathrm{p}}\xspace}                 

 \def\Ps      {\ensuremath{\mathrm{s}}\xspace}

 \def\thebaroffset{0.0em}
}
{

 \def\Ppi         {\ensuremath{\pi}\xspace}

 \def\Pchi        {\ensuremath{\chi}\xspace}

 \mathchardef\PDelta="7101
 \mathchardef\PXi="7104
 \mathchardef\PLambda="7103
 \mathchardef\PSigma="7106
 \mathchardef\POmega="710A
 \mathchardef\PUpsilon="7107
 \mathchardef\PPi="7105
 \def\PB      {\ensuremath{B}\xspace}                 
 \def\PD      {\ensuremath{D}\xspace}                 

 \def\PK      {\ensuremath{K}\xspace}                 
 \def\Pb      {\ensuremath{b}\xspace}                 
 \def\Pc      {\ensuremath{c}\xspace}

 \def\Pp      {\ensuremath{p}\xspace}                 

 \def\Ps      {\ensuremath{s}\xspace}

 \def\thebaroffset{0.18em}
}
\newcommand{\offsetoverline}[2][\thebaroffset]{\kern #1\overline{\kern -#1 #2}}%

\makeatletter
\ifcase \@ptsize \relax% 10pt
  \newcommand{\miniscule}{\@setfontsize\miniscule{4}{5}}% \tiny: 5/6
\or% 11pt
  \newcommand{\miniscule}{\@setfontsize\miniscule{5}{6}}% \tiny: 6/7
\or% 12pt
  \newcommand{\miniscule}{\@setfontsize\miniscule{5}{6}}% \tiny: 6/7
\fi
\makeatother

\DeclareRobustCommand{\optbar}[1]{\shortstack{{\miniscule (\rule[.5ex]{1.25em}{.18mm})}
  \\ [-.7ex] $#1$}}

\def\squark    {{\ensuremath{\Ps}}\xspace}

\def\cquark    {{\ensuremath{\Pc}}\xspace}
\def\cquarkbar {{\ensuremath{\overline \cquark}}\xspace}

\def\bquark    {{\ensuremath{\Pb}}\xspace}

\def\pion   {{\ensuremath{\Ppi}}\xspace}

\def\pip    {{\ensuremath{\pion^+}}\xspace}
\def\pim    {{\ensuremath{\pion^-}}\xspace}

\def\kaon    {{\ensuremath{\PK}}\xspace}
\def\KorKbar {\kern \thebaroffset\optbar{\kern -\thebaroffset \PK}{}\xspace}

\def\Kp      {{\ensuremath{\kaon^+}}\xspace}
\def\Km      {{\ensuremath{\kaon^-}}\xspace}

\def\KS      {{\ensuremath{\kaon^0_{\mathrm{S}}}}\xspace}

\def\D       {{\ensuremath{\PD}}\xspace}

\def\DorDbar {\kern \thebaroffset\optbar{\kern -\thebaroffset \PD}\xspace}
\def\Dz      {{\ensuremath{\D^0}}\xspace}

\def\Dp      {{\ensuremath{\D^+}}\xspace}
\def\Dm      {{\ensuremath{\D^-}}\xspace}

\def\DpDm    {\ensuremath{\Dp {\kern -0.16em \Dm}}\xspace}

\def\B       {{\ensuremath{\PB}}\xspace}

\def\BorBbar {\kern \thebaroffset\optbar{\kern -\thebaroffset \PB}\xspace}

\def\Bd      {{\ensuremath{\B^0}}\xspace}

\def\BdorBdbar {\kern \thebaroffset\optbar{\kern -\thebaroffset \Bd}\xspace}
\def\Bu      {{\ensuremath{\B^+}}\xspace}

\def\Bs      {{\ensuremath{\B^0_\squark}}\xspace}

\def\BsorBsbar {\kern \thebaroffset\optbar{\kern -\thebaroffset \Bs}\xspace}

\def\Bds     {{\ensuremath{\B_{(\squark)}^0}}\xspace}

\def\chiczero {{\ensuremath{\Pchi_{\cquark 0}}}\xspace}

\def\Y#1S{\ensuremath{\PUpsilon{(#1S)}}\xspace}

\def\proton      {{\ensuremath{\Pp}}\xspace}
\def\antiproton  {{\ensuremath{\overline \proton}}\xspace}

\def\Lz          {{\ensuremath{\PLambda}}\xspace}
\def\Lbar        {{\ensuremath{\offsetoverline{\PLambda}}}\xspace}
\def\LorLbar     {\kern \thebaroffset\optbar{\kern -\thebaroffset \PLambda}\xspace}

\def\Sigmares    {{\ensuremath{\PSigma}}\xspace}
\def\Sigmaz      {{\ensuremath{\Sigmares{}^0}}\xspace}

\def\Xires       {{\ensuremath{\PXi}}\xspace}

\def\Lc          {{\ensuremath{\Lz^+_\cquark}}\xspace}
\def\Lcbar       {{\ensuremath{\Lbar{}^-_\cquark}}\xspace}

\def\Sigmacz      {{\ensuremath{\Sigmares_\cquark^0}}\xspace}

\def\Xicp        {{\ensuremath{\Xires^+_\cquark}}\xspace}

\def\Lb           {{\ensuremath{\Lz^0_\bquark}}\xspace}

\def\Xib          {{\ensuremath{\Xires_\bquark}}\xspace}
\def\Xibz         {{\ensuremath{\Xires^0_\bquark}}\xspace}

\def\BF         {{\ensuremath{\mathcal{B}}}\xspace}

\newcommand{\decay}[2]{\ensuremath{\mathinner{#1\!\to #2}}\xspace}

\def\to                 {\ensuremath{\rightarrow}\xspace}

\def\CP                {{\ensuremath{C\!P}}\xspace}

\def\AT#1     {\ensuremath{A_{\mathrm{T}}^{#1}}\xspace}           % 2

\def\C#1      {\ensuremath{\mathcal{C}_{#1}}\xspace}                       % 9
\def\Cp#1     {\ensuremath{\mathcal{C}_{#1}^{'}}\xspace}                    % 7
\def\Ceff#1   {\ensuremath{\mathcal{C}_{#1}^{\mathrm{(eff)}}}\xspace}        % 9  
\def\Cpeff#1  {\ensuremath{\mathcal{C}_{#1}^{'\mathrm{(eff)}}}\xspace}       % 7
\def\Ope#1    {\ensuremath{\mathcal{O}_{#1}}\xspace}                       % 2
\def\Opep#1   {\ensuremath{\mathcal{O}_{#1}^{'}}\xspace}                    % 7

\newcommand{\nospaceunit}[1]{\ensuremath{\text{#1}}}       
\newcommand{\aunit}[1]{\ensuremath{\text{\,#1}}}       
\newcommand{\tev}{\aunit{Te\kern -0.1em V}\xspace}
\newcommand{\gev}{\aunit{Ge\kern -0.1em V}\xspace}
\newcommand{\mev}{\aunit{Me\kern -0.1em V}\xspace}
\newcommand{\kev}{\aunit{ke\kern -0.1em V}\xspace}
\newcommand{\ev}{\aunit{e\kern -0.1em V}\xspace}
 
\newcommand{\mevc}{\ensuremath{\aunit{Me\kern -0.1em V\!/}c}\xspace}
\newcommand{\gevc}{\ensuremath{\aunit{Ge\kern -0.1em V\!/}c}\xspace}
\newcommand{\mevcc}{\ensuremath{\aunit{Me\kern -0.1em V\!/}c^2}\xspace}
\newcommand{\gevcc}{\ensuremath{\aunit{Ge\kern -0.1em V\!/}c^2}\xspace}
\def\mum  {\ensuremath{\,\upmu\nospaceunit{m}}\xspace}

\def\fb   {\ensuremath{\aunit{fb}}\xspace}
\def\invfb   {\ensuremath{\fb^{-1}}\xspace}

\newcommand{\stat}{\aunit{(stat)}\xspace}
\newcommand{\syst}{\aunit{(syst)}\xspace}

\def\gsim{{~\raise.15em\hbox{$>$}\kern-.85em
          \lower.35em\hbox{$\sim$}~}\xspace}
\def\lsim{{~\raise.15em\hbox{$<$}\kern-.85em
          \lower.35em\hbox{$\sim$}~}\xspace}

\def\sPlot{\mbox{\em sPlot}\xspace}

\def\pt         {\ensuremath{p_{\mathrm{T}}}\xspace}

\def\ptot       {\ensuremath{p}\xspace}

\def\evtgen     {\mbox{\textsc{EvtGen}}\xspace}

\def\geant      {\mbox{\textsc{Geant4}}\xspace}

\def\photos     {\mbox{\textsc{Photos}}\xspace}

\def\pythia     {\mbox{\textsc{Pythia}}\xspace}

\def\roofit     {\mbox{\textsc{RooFit}}\xspace}

\def\tell1  {TELL1\xspace}
\def\ukl1   {UKL1\xspace}

\newcommand{\lhcborcid}[1]{\href{https://orcid.org/#1}{\hspace*{0.1em}\raisebox{-0.45ex}{\includegraphics[width=1em]{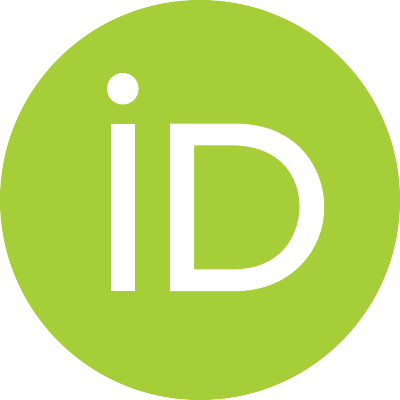}}}}

\hypersetup{
  colorlinks   = true, %Colours links instead of ugly boxes
  urlcolor     = blue, %Colour for external hyperlinks
  linkcolor    = blue, %Colour of internal links
  citecolor    = red   %Colour of citations
}

\ifEnableSectionTOCLinks
    \usepackage[explicit]{titlesec} % to change headings
    
    \let\oldcontentsline\contentsline
    \renewcommand

    \titleformat{\section}{\normalfont\Large\bf}{\hyperlink{tocsection.\thesection}{{\thesection} \parbox[t]{\dimexpr\textwidth-1pc}{#1}}}{1pc}{}

    \titleformat{\subsection}{\normalfont\bf}{\hyperlink{tocsubsection.\thesubsection}{{\thesubsection} \parbox[t]{\dimexpr\textwidth-1pc}{#1}}}{1pc}{}

    \titleformat{name=\section,numberless}[display]{}{}{0pt}{\normalfont\Huge\bfseries #1}
\fi

\usepackage{cite} % Allows for ranges in citations
\usepackage{LHCb/mciteplus}
\def\Sz{{\ensuremath{\Sigmares{}^0}}\xspace}

\def\Bds{{\ensuremath{\B^0_{\kern -0.1em{\scriptscriptstyle (}\kern -0.05em\squark\kern -0.03em{\scriptscriptstyle )}}}}\xspace}

\newcommand{\KK}{\Kp \Km}
\newcommand{\PPbar}{\proton \antiproton}

\newcommand{\LPPbar}{\Lz \proton \antiproton}

\newcommand{\LPPim}{\texorpdfstring{\decay{\Lz}{\proton\pim}}{}}

\newcommand{\BdPPbar}{\texorpdfstring{\decay{\Bd}{\proton \antiproton}}{}}

\newcommand{\BPPbarHH}{\texorpdfstring{\decay{\Bds}{\proton \antiproton h^+ h^{\prime-}}}{}}
\newcommand{\BdsLPH}{\texorpdfstring{\decay{\Bds}{\Lz \proton h^-}}{}}
\newcommand{\BdPPbarPPbar}{\texorpdfstring{\decay{\Bd}{\proton \antiproton \proton \antiproton}}{}}
\newcommand{\LbLChiczOneP}{\texorpdfstring{\decay{\Lb}{\Lz \chiczero(1P)}}{}}
\newcommand{\LbLHH}{\texorpdfstring{\decay{\Lb}{\Lz h^+ h^{\prime-}}}{}}

\newcommand{\LbLKK}{\texorpdfstring{\decay{\Lb}{\Lz \Kp \Km}}{}}

\newcommand{\LbLPPbar}{\texorpdfstring{\decay{\Lb}{\Lz \proton \antiproton}}{}}

\newcommand{\LbSzKK}{\texorpdfstring{\decay{\Lb}{\Sz \Kp \Km}}{}}

\def\LbToLcpppi  {\decay{\Lb}{\Lc \proton \antiproton \pim}}
\newcommand{\XibzLPPbar}{\texorpdfstring{\decay{\Xibz}{\Lz \proton \antiproton}}{}}

\def\BsToLcLc  {\decay{\Bs}{\Lc \Lcbar}}
\def\BuToLPPbarP  {\decay{\Bu}{\Lbar \proton \antiproton \proton}}

\def\Sigmacz     {{\ensuremath{\Sigmares{}^0_\cquark}}\xspace}
\def\Sigmacstarz     {{\ensuremath{\Sigmares{}^{*0}_\cquark}}\xspace}
\def\LbToScPPbar  {\decay{\Lb}{\Sigmacz \proton \antiproton}}
\def\LbToScstPPbar  {\decay{\Lb}{\Sigmacstarz \proton \antiproton}}
\usepackage{longtable} % only for template; not usually to be used in PAPERs
\usepackage{comment}

\usepackage[compat=1.1.0]{tikz-feynman}

\begin{document}

%%%%%%%%%%%%%%%%%%%%%%%%%
%%%%% Title     %%%%%%%%%
%%%%%%%%%%%%%%%%%%%%%%%%%
\renewcommand{\thefootnote}{\fnsymbol{footnote}}
\setcounter{footnote}{1}

% %%%%%%% CHOOSE TITLE PAGE--------
% ===============================================================================
% Purpose: LHCb-PAPER journal paper title page template
% Author: 
% Created on: 2010-09-25
% ===============================================================================

%%%%%%%%%%%%%%%%%%%%%%%%%
%%%%%  TITLE PAGE  %%%%%%
%%%%%%%%%%%%%%%%%%%%%%%%%
\begin{titlepage}
\pagenumbering{roman}

% Header ---------------------------------------------------
\vspace*{-1.5cm}
\centerline{\large EUROPEAN ORGANIZATION FOR NUCLEAR RESEARCH (CERN)}
\vspace*{1.5cm}
\noindent
\begin{tabular*}{\linewidth}{lc@{\extracolsep{\fill}}r@{\extracolsep{0pt}}}
\ifthenelse{\boolean{pdflatex}}% Logo format choice
{\vspace*{-1.5cm}\mbox{\!\!\!\includegraphics[width=.14\textwidth]{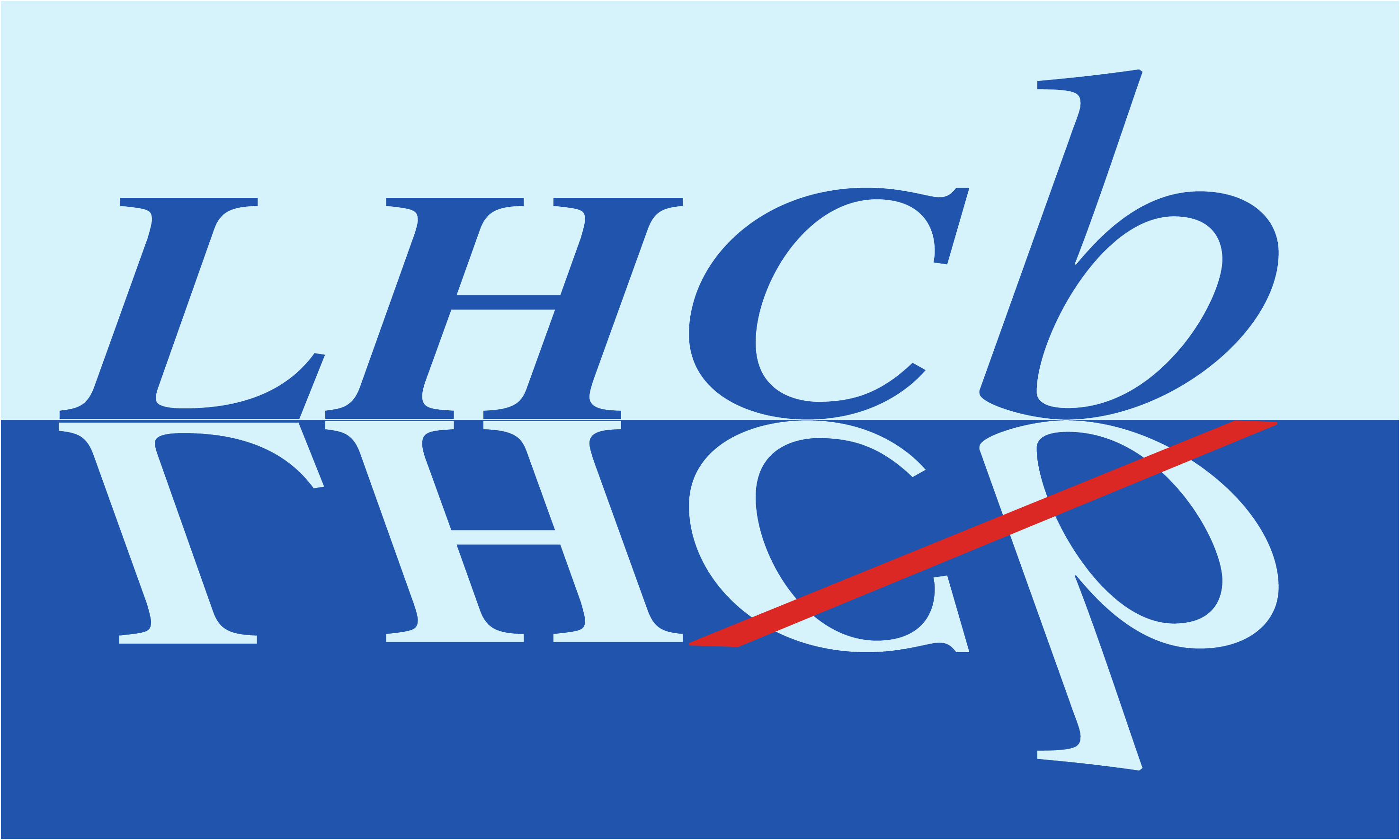}} & &}%
{\vspace*{-1.2cm}\mbox{\!\!\!\includegraphics[width=.12\textwidth]{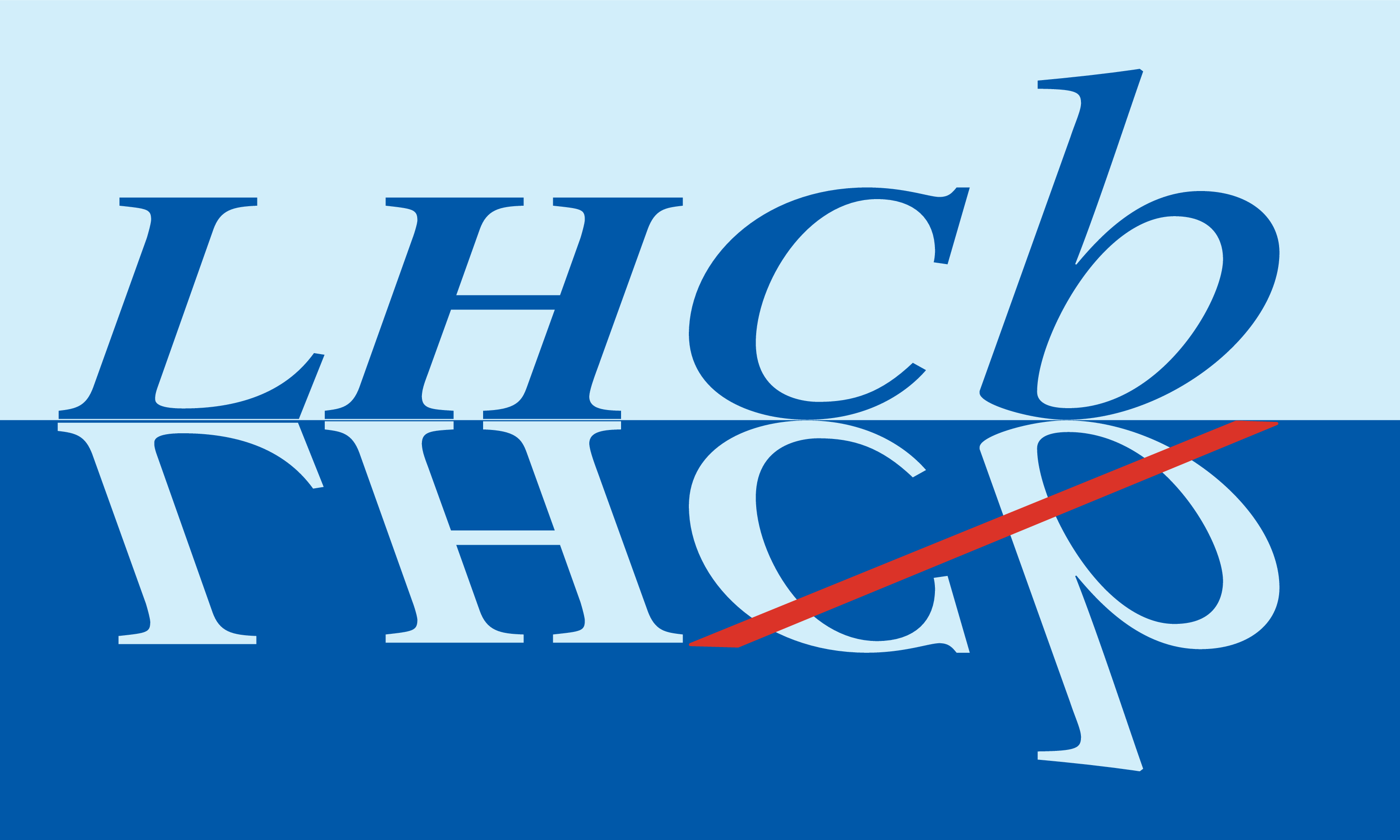}} & &}%
\\
 & & CERN-EP-2026-123 \\  % ID 
 & & LHCb-PAPER-2026-004 \\  % ID 
 & & May 4, 2026 \\ % Date - Can also hardwire e.g.: 23 March 2010
 & & \\
% not in paper \hline
\end{tabular*}

\vspace*{4.0cm}

% Title --------------------------------------------------
{\normalfont\bfseries\boldmath\huge
\begin{center}
% DO NOT EDIT HERE. Instead edit macro in main.tex to keep metadata correct
  \papertitle 
\end{center}
}

\vspace*{2.0cm}

% Authors -------------------------------------------------
\begin{center}
% Edit macro in main.tex to keep metadata correct
\paperauthors\footnote{Authors are listed at the end of this paper.}
\end{center}

\vspace{\fill}

% Abstract -----------------------------------------------
\begin{abstract}
  \noindent
A search for the charmless purely baryonic decay $\mathinner{\mathit{\Lambda}^0_b\!\to \mathit{\Lambda} p \overline{p}}$ is performed using proton-proton collision data recorded by the LHCb experiment at a centre-of-mass energy of $\sqrt{s}=13\,\text{TeV}$ and corresponding to an integrated luminosity of $6.0\,\text{fb}^{-1}$. The signal decay is observed with a significance of 5.1 standard deviations. Its branching fraction is measured for the first time, relative to that of the topologically similar decay $\mathinner{\mathit{\Lambda}^0_b\!\to \mathit{\Lambda} K^+ K^-}$. Contributions from intermediate charmonium resonances decaying to the $p \overline{p}$ and $K^+ K^-$ final states are explicitly excluded with a requirement on the invariant mass of the companion hadron system, $m(h\bar{h}) < 2.85\,\text{GeV}$, where $h$ stands for a proton or a charged kaon.
The relative branching fraction is found to be
$$
\frac{\mathcal{B}(\mathinner{\mathit{\Lambda}^0_b\!\to \mathit{\Lambda} p \overline{p}})}{\mathcal{B}(\mathinner{\mathit{\Lambda}^0_b\!\to \mathit{\Lambda} K^+ K^-})} = (5.1 \pm 1.3_{\stat} \pm 0.3_{\syst}) \times 10^{-2} \,.
$$
\end{abstract}

\vspace*{2.0cm}

\begin{center}
Published in Journal of High Energy Physics 08 (2026) 040
\end{center}

\vspace{\fill}

{\footnotesize 
% Edit macro in main.tex to keep metadata correct
\centerline{\copyright~\papercopyright. \href{\paperlicenceurl}{\paperlicence}.}}
\vspace*{2mm}

\end{titlepage}

%%%%%%%%%%%%%%%%%%%%%%%%%%%%%%%%
%%%%%  EOD OF TITLE PAGE  %%%%%%
%%%%%%%%%%%%%%%%%%%%%%%%%%%%%%%%

%  empty page follows the title page ----
\newpage
\setcounter{page}{2}
\mbox{~}

%\twocolumn
% %%%%%%%%%%%%% ---------

\renewcommand{\thefootnote}{\arabic{footnote}}
\setcounter{footnote}{0}

%%%%%%%%%%%%%%%%%%%%%%%%%%%%%%%%
%%%%%  Table of Content   %%%%%%
%%%%%%%%%%%%%%%%%%%%%%%%%%%%%%%%
%%%% Uncomment if desired
%\tableofcontents

\cleardoublepage

%%%%%%%%%%%%%%%%%%%%%%%%%
%%%%% Main text %%%%%%%%%
%%%%%%%%%%%%%%%%%%%%%%%%%

\pagestyle{plain} % restore page numbers for the main text
\setcounter{page}{1}
\pagenumbering{arabic}

%% Uncomment during review phase. 
%% Comment before a final submission.
%\linenumbers

%% This is the main body
%% It is useful to have a single file so comments are not missed in overleaf.
\section{Introduction}
\label{sec:Introduction}
The study of decays of \bquark hadrons to baryonic final states is an active research field in hadronic flavour physics, with several new decay modes observed in recent years~\cite{PDG2024}. In particular, the \lhcb collaboration observed the first examples of charmless four-body baryonic $B$ decays in 2016, namely the \BPPbarHH modes~\cite{LHCb-PAPER-2017-005}.\footnote{Here, $h$ and $h'$ denote either a charged pion or a charged kaon.}
This was followed in 2017 by the report of the first decay of a $B$ meson into a purely baryonic final state, \BdPPbar~\cite{LHCb-PAPER-2017-022}.
Searches for both more experimentally challenging and theoretically unexplored baryonic final states have followed, and again the \lhcb collaboration published the first observations of the \BdPPbarPPbar~\cite{LHCb-PAPER-2022-032}, \BuToLPPbarP~\cite{LHCb-PAPER-2025-032} and \BsToLcLc~\cite{LHCb-PAPER-2025-053} decays.

The Standard Model predicts the existence of many decays to baryonic final states, including the so-called \textit{purely baryonic decay processes}~\cite{PBD1,PBD2,Geng:2026cyx} that only involve baryons. Yet, they remain largely unexplored. Currently, the only purely baryonic decay modes measured are \LbToScPPbar and \LbToScstPPbar, observed by \lhcb in the study of the decay mode \LbToLcpppi in 2018~\cite{LHCb-PAPER-2018-005}.
Purely baryonic decays can be exploited to measure \CP asymmetries or to study time-reversal ($T$) symmetry violations with triple-product correlations.
The latter are especially interesting in purely baryonic decays owing to their rich spin structures~\cite{PBD1,PBD2,Geng:2026cyx}, complementing measurements performed with mesonic final states. Predictions for direct \CP asymmetries of approximately 3\% and $-13\%$ exist for the decay modes \LbLPPbar and \XibzLPPbar, respectively~\cite{PBD1}.
Similarly to what is found in the study of baryonic $b$-hadron decays, enhancements at the mass threshold of final-state di-baryon pairs are to be expected also for purely baryonic decays~\cite{PBD1,PBD2,Geng:2026cyx}.

\begin{figure}[!tp]
        \begin{center}
           \includegraphics[width=0.45\textwidth]{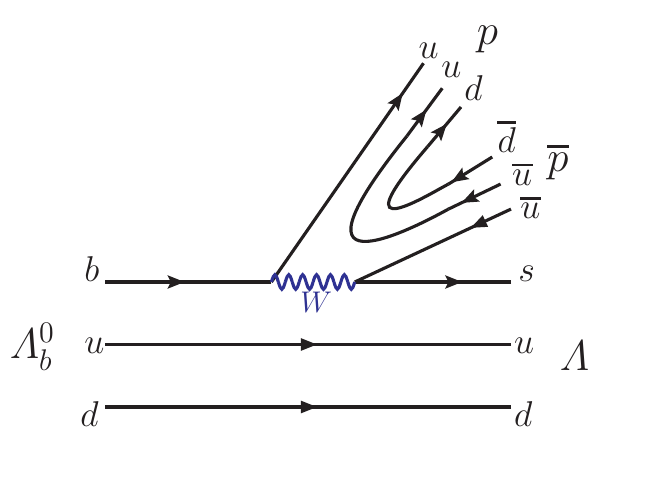}
           \includegraphics[width=0.45\textwidth]{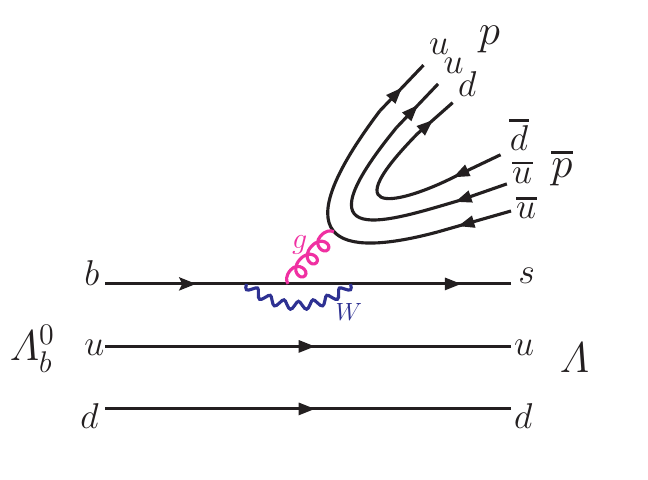}\\
           \includegraphics[width=0.45\textwidth]{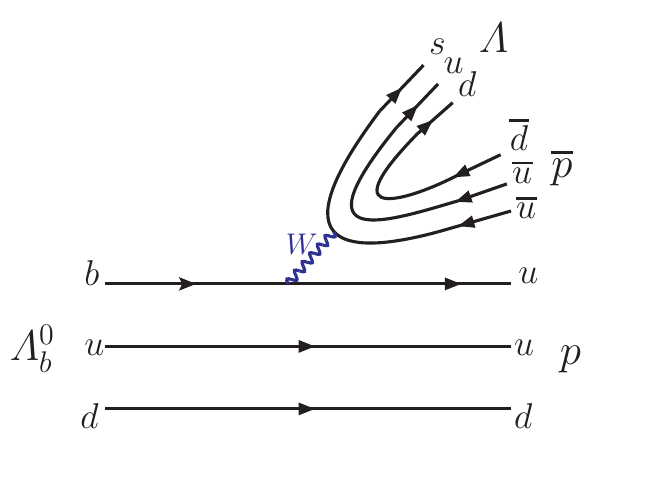}
           \includegraphics[width=0.45\textwidth]{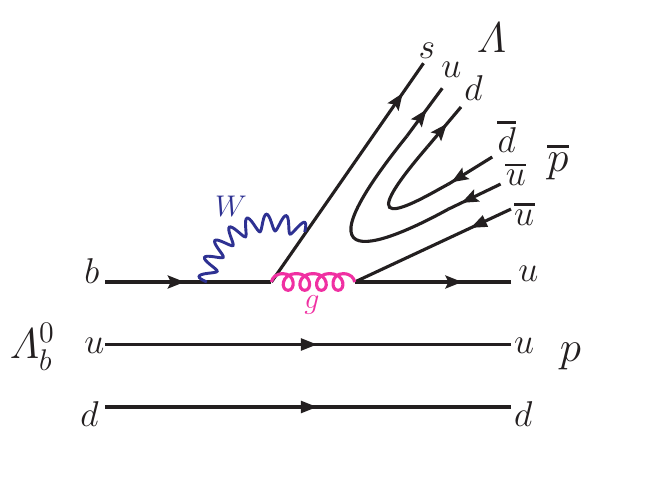}
        \end{center}
    \caption{Feynman diagrams describing the purely baryonic decay $\mathinner{\mathit{\Lambda}^0_b\!\to \mathit{\Lambda} p \overline{p}}$.}
        \label{Feynman_diagrams}
    \end{figure}

This paper presents the first search for the charmless \LbLPPbar decay mode, and the measurement of its branching fraction relative to that of the topologically similar decay \LbLKK~\cite{LHCb-PAPER-2016-004}.
The decay \LbLPPbar is mediated by the dominant Feynman diagrams presented in Fig.~\ref{Feynman_diagrams}, from which
theoretical calculations, evaluated on the full phase space, predict the branching fraction
$\mathcal{B}(\LbLPPbar) = (3.2^{+0.8}_{-0.3} \pm 0.4 \pm 0.7) \times 10^{-6}$~\cite{PBD1},
where the uncertainties are associated with nonfactorisable effects, CKM matrix elements, and hadronic form factors, respectively.

\section{Detector and simulation}
\label{sec:DetectorAndSim}
The \lhcb detector~\cite{LHCb-DP-2008-001,LHCb-DP-2014-002} is a single-arm forward
spectrometer covering the \mbox{pseudorapidity} range $2<\eta <5$,
designed for the study of particles containing \bquark or \cquark
quarks. The detector components used to collect the data analysed in this paper include a high-precision tracking system
consisting of a silicon-strip vertex detector surrounding the $pp$
interaction region~\cite{LHCb-DP-2014-001}, a large-area silicon-strip detector located
upstream of a dipole magnet with a bending power of about
$4{\mathrm{\,T\,m}}$, and three stations of silicon-strip detectors and straw
drift tubes~\cite{LHCb-DP-2017-001}
placed downstream of the magnet.
The tracking system provides a measurement of the momentum, \ptot, of charged particles with
a relative uncertainty that varies from 0.5\% at low momentum to 1.0\% at 200\gev.\footnote{Natural units with $\hbar = c = 1$ are used throughout.}
The minimum distance of a track to a primary $pp$ collision vertex (PV), the impact parameter (IP), 
is measured with a resolution of $(15+29/\pt)\mum$,
where \pt is the component of the momentum transverse to the beam, in \gev.
Different types of charged hadrons are distinguished using information
from two ring-imaging Cherenkov (RICH) detectors~\cite{LHCb-DP-2012-003}. 
Photons, electrons and hadrons are identified by a calorimeter system consisting of
scintillating-pad and preshower detectors, an electromagnetic 
and a hadronic calorimeter. Muons are identified by a
system composed of alternating layers of iron and multiwire
proportional chambers~\cite{LHCb-DP-2012-002}.
The online event selection is performed by a trigger~\cite{LHCb-DP-2012-004}, 
which consists of a hardware stage, based on information from the calorimeter and muon
systems, followed by a software stage, which applies a full event
reconstruction.
Triggered data further undergo a centralised, offline processing step
to deliver physics-analysis-ready data across the entire \lhcb physics programme~\cite{Stripping}.

At the hardware trigger stage, events must either contain a hadron candidate
with sufficiently large transverse energy, or be triggered independently of the signal candidate by particles originating from the rest of the event. In the software trigger, at least one high-quality track with
large \pt and significant IP with respect
to any PV is required. The selection further demands that
two- or three-track combinations form a secondary vertex that is
significantly displaced from all PVs. Boosted-decision-tree algorithms are also used to
identify vertices consistent with the decay of a $b$ hadron. The same trigger
requirements are applied to both the \LbLPPbar signal candidates and the
\LbLKK normalisation mode.

Simulation is required to model the effects of the detector acceptance and the imposed selection requirements. In the simulation, $pp$ collisions are generated using \pythia~\cite{Sjostrand:2007gs,*Sjostrand:2006za} with a specific \lhcb configuration~\cite{LHCb-PROC-2010-056}. Decays of unstable particles are described by \evtgen~\cite{Lange:2001uf}, in which final-state radiation is generated using \photos~\cite{davidson2015photos}. The interaction of the generated particles with the detector, and its response, are implemented using the \geant toolkit~\cite{Allison:2006ve, *Agostinelli:2002hh} as described in Ref.~\cite{LHCb-PROC-2011-006}. Simulated samples are produced for Run~2 data-taking conditions and processed through the full detector simulation, trigger emulation, and reconstruction chains. These samples are used to determine efficiencies, model the signal and partially reconstructed components entering the fits, study background contributions, and validate the analysis procedure.
The ROOT~\cite{Root} and \lhcb~\cite{KOPPENBURG2006213,Tsaregorodtsev:2010zz} software frameworks are used for the initial data preparation, while the statistical analysis is performed with the \roofit library~\cite{roofit}.

The simulated three-body decays are generated uniformly in the \emph{square Dalitz} variables~\cite{PhysRevD.72.052002}. These induce a transformation that makes the original Dalitz plot square, in a way that the most populated regions are extended, providing a more uniform coverage of the kinematic region to study acceptance and efficiencies.

\section{Analysis overview and dataset}
\label{sec:analysis-strategy}

The analysis is based on the full Run~2 dataset of $pp$ collisions collected by the \lhcb experiment at $\sqrt{s}=13\tev$ during the 2015--2018 data-taking period, corresponding to an
integrated luminosity of 6.0\invfb. The measurement determines the ratio between the branching fractions of the \LbLPPbar and the
topologically similar \LbLKK decays. Using a
normalisation mode with the same parent particle and a closely related
final state allows several systematic effects to cancel or be minimised in the ratio. In particular, the uncertainties related to the \Lb production cross-section and the
integrated luminosity cancel, while differences in the
reconstruction and selection efficiencies are significantly reduced. Given the similarity between the normalisation mode and the signal, an identical selection strategy is applied to both channels and the same categorisation is used.

The branching-fraction ratio is obtained from
\begin{equation}
R_{\Lb}\equiv\frac{\BF(\LbLPPbar)}
     {\BF(\LbLKK)}
=
\frac{N(\LbLPPbar)}
     {N(\LbLKK)}
\cdot
\frac{\epsilon_{\LbLKK}}
     {\epsilon_{\LbLPPbar}} \,,
\label{eq:bf_ratio}
\end{equation}
\noindent
where $N$ denotes the fitted signal yields and $\epsilon$ the respective total
efficiencies, including detector acceptance, trigger, reconstruction and selection effects.

The \Lz baryons are reconstructed via the decay mode \LPPim in two different categories: the first involving \Lz baryons that decay early enough for the pion and proton to be reconstructed in the vertex detector;
and the second containing \Lz baryons that decay outside of this detector, such that track segments cannot be formed within its acceptance. These track pairs are categorized as \emph{long--long} (LL) and \emph{downstream--downstream} (DD), respectively. The LL category has better mass, momentum and vertex resolution than the DD category. Nevertheless, about half of the observed \LbLPPbar candidates are reconstructed in the DD category, so it is retained in the analysis. The LL and DD \Lz baryon categories are treated separately in the data and simulation processing, yielding a total of eight independent subsamples corresponding to the four data-taking years and the two track categories.

The contribution from the \XibzLPPbar decay, which shares the same final state as the signal channel, is considered.
Although this decay is not expected to be observed, since its predicted branching fraction is about one order of magnitude smaller than that of the decay under study~\cite{PBD1}, its possible contribution, 
located 175\mev above the \Lb signal peak,  is explicitly included in the fit used to determine the signal yield.

A simultaneous unbinned maximum-likelihood fit to the invariant-mass
distributions of the signal and normalisation candidates is performed. To avoid experimenter's bias, candidates with
reconstructed \LPPbar invariant-mass within $\pm 50\mev$\ of the known \Lb and
\Xib masses were initially excluded until the event selection, fit model, and systematic
studies had been finalised. Sideband data, simulated samples, and the normalisation
mode are used to develop and validate the analysis procedure. The signal significance and branching-fraction ratio are evaluated only after the fit model and its validation are finalised.

\section{Selection}
\label{sec:Selection}

\subsection{Preliminary selection}
After triggering, a preliminary selection is applied to fully reconstruct and select the signal and normalisation decay chains.
The proton and pion tracks from the \LPPim decay are
required to originate from a common vertex, to have reconstruction quality and a significant IP with respect to any PV. In addition, both tracks in the LL category are required to satisfy $\pt > 250\mev$. Loose particle-identification (PID) requirements are applied to the
proton in the LL category, providing effective suppression of
misidentified \Lz candidates, while no PID requirements are imposed in the
DD category at this stage. Additional requirements are applied to the reconstructed \Lz candidate on the position of its decay vertex, its separation from the \Lb decay vertex, and its reconstructed mass.

The reconstructed \Lz candidate is then combined with either a \PPbar
or a \KK pair to form a \Lb candidate. The companion hadrons must satisfy track quality, \pt and
IP requirements, and appropriate PID selections are applied
to distinguish protons from kaons. The topology and kinematics of the
reconstructed decay, including the quality and displacement of the \Lb decay
vertex from the PV, are used to suppress backgrounds.
Apart from the PID applied to the
companion hadrons, identical reconstruction and selection requirements are
used for both the \LbLPPbar and \LbLKK decay modes to ensure a close cancellation of systematic
effects in the branching-fraction ratio.

Candidates consistent with \Lb baryons are restricted to the mass range
\mbox{$5350 < m(\Lz h \bar{h}) < 6050\mev$}.\footnote{Unless specified otherwise, $h$ denotes a proton or a charged kaon throughout the paper.} Simulated \LbLPPbar and \LbLKK samples are used to model the signal distributions of the variables entering the selection optimisation. The background is modelled using data from the upper sideband of the $m(\LPPbar)$ spectrum, defined by the intervals $5669 < m(\LPPbar) < 5738\mev$ and $5838 < m(\LPPbar) < 6100\mev$, excluding the \Xibz region. The lower sideband is not used due to possible contamination from partially reconstructed decays.

The  kinematic quantities are refined by performing, separately for the signal and normalisation samples, a fit to the full \decay{\Lb}{\Lz h \bar{h}} decay chain~\cite{Hulsbergen:2005pu}, in which the decay products are constrained to a common vertex, the \Lz mass is constrained to its known value~\cite{PDG2024}, and the
\Lb candidate is constrained to originate from the associated PV.
Kinematic variables obtained from this constrained fit are used alongside their
counterparts computed without it in the subsequent analysis.

In the simulation, a truth-matching requirement is applied after the preselection to retain only candidates whose reconstructed final-state particles
can be matched to the generated decay chain.
The fraction of candidates failing this requirement is found to differ between the signal and normalisation
modes. Its possible impact on the efficiency ratio and on the measured
branching-fraction ratio is therefore evaluated separately and accounted for with a systematic uncertainty.

\subsection{Multivariate and PID selections}\label{sec:MVA_PID}

A multivariate classifier (\textsc{XGBoost} algorithm~\cite{XGBoosting} implemented in the Python library of Ref.~\cite{xgboost_GitHub}) is used to enhance the separation between signal
decays and combinatorial background beyond what is achievable with the
preselection alone. The classifier is trained using simulated
\LbLKK decays to represent the signal, whereas background candidates are taken from the upper sideband of the invariant-mass
distribution in data, as defined earlier. The LL and DD categories are treated
independently, yielding two separately optimised classifiers.  An
alternative classifier trained with simulated \LbLPPbar signal decays was also
considered. Its performance, assessed from the corresponding receiver-operating-characteristic
curves, was found to be consistent with that of the classifier trained on \LbLKK decays within the statistical precision of the simulated samples, and the latter is therefore
used for both channels.

The set of input variables used to train the classifier is chosen based on their separation power and overall classifier performance. These variables
describe the kinematic and topological properties of the \Lb candidate, including its \pt and $\eta$, the alignment between the reconstructed momentum and the flight direction, as well as the quality and displacement of the decay vertices of the \Lb and \Lz candidates. Studies performed on upper-sideband
background candidates show no significant correlation between the classifier response and the reconstructed invariant mass. In addition, the classifier
response is compared between training and testing samples using
Kolmogorov--Smirnov tests, which show no evidence of overtraining. The  classifier output is then used as one of the inputs to the final event selection.

Particle-identification variables are used to distinguish protons, kaons and pions in
the final state using information from the \lhcb 
detectors, in particular the RICH system, together with track-quality and event-related
variables. Machine learning algorithms combine these inputs to provide,
for each track, probabilities for the different particle hypotheses~\cite{LHCb-DP-2018-001}.
For the \LbLKK normalisation mode, a
combined PID variable is constructed from the per-track kaon, pion and
proton probabilities to select genuine kaon pairs and suppress
misidentified backgrounds. An analogous variable is defined for the proton and antiproton in the \LbLPPbar signal mode.

The same \textsc{XGBoost} classifier response is used for both the signal and normalisation modes. In each mode, the requirement on this response is optimised jointly with that on the corresponding combined PID variable to maximise the sensitivity to the \LbLPPbar signal while retaining a robust selection for the normalisation mode. The baseline selection is determined through a two-dimensional optimisation based on the Punzi figure of merit~\cite{Punzi:2003bu} $\varepsilon_{\mathrm{s}}/(a/2+\sqrt{N_B})$, where $\varepsilon_{\mathrm{s}}$ is the signal efficiency, $N_B$ is the expected combinatorial background yield in the signal region, and $a=5$ corresponds to the targeted significance in units of standard deviations. The optimisation is performed independently for the LL and DD categories.

For each set of MVA and PID requirements, the signal efficiency $\varepsilon_S$
is determined from truth-matched simulated signal candidates, while the expected
combinatorial background $N_B$ is evaluated in a mass window centred on the known
\Lb mass and defined as $\pm 1.5\sigma_m$, where
$\sigma_m = 12\mev$ is the mass resolution estimated from simulation.
The expected combinatorial background in this window is obtained fitting the upper sideband of the invariant-mass distribution with a linear function and extrapolating this fit result to the signal region. This
procedure yields stable and well-defined optimal requirements for both
decay modes. Several alternative optimisation strategies were investigated and found to give consistent results but no improvement in performance.

\renewcommand{\arraystretch}{1.6}
\begin{table}[b]
\centering
\caption{Charm-hadron vetoes applied to the $\mathit{\Lambda}^0_b$ candidates for the two decays under study. 
The notation $m(K^\pm\pi^\mp)_{KK}$ denotes the invariant mass of the $K^+ K^-$ pair computed after assigning the pion mass to either of the two kaon tracks (\textit{i.e.}, considering both $K^+ \pi^-$ and $\pi^+ K^-$ hypotheses).}
\label{tab:vetoes}
\begin{tabular}{l l}
\hline\hline
\multicolumn{2}{c}{\rule{0mm}{5mm}\LbLKK} \\[0mm]
\hline
$\left| m(\KK) - m_{\Dz} \right|$                      & $> 30\mev$ \\
$\left| m(\Lz \Kp) - m_{\Lc} \right|$         & $> 30\mev$ \\
$\left| m(\Lz \Kp) - m_{\Xicp} \right|$             & $> 30\mev$ \\
$\left| m(K^\pm \pi^\mp)_{KK} - m_{\Dz} \right|$      & $> 30\mev$ \\
$m(\KK)$                                               & $< 2.85\gev$ \\
\hline
\multicolumn{2}{c}{\rule{0mm}{5mm}\LbLPPbar}\\[0mm]
\hline
$m(\PPbar)$ & $< 2.85\gev$ \\
\hline\hline
\end{tabular}
\end{table}
\renewcommand{\arraystretch}{1.0}

\subsection{Charm vetoes, multiple candidates and background study}
\label{sec:vetos}

\begin{figure}[t]
    \centering
    % Top row
    \includegraphics[width=0.42\textwidth]{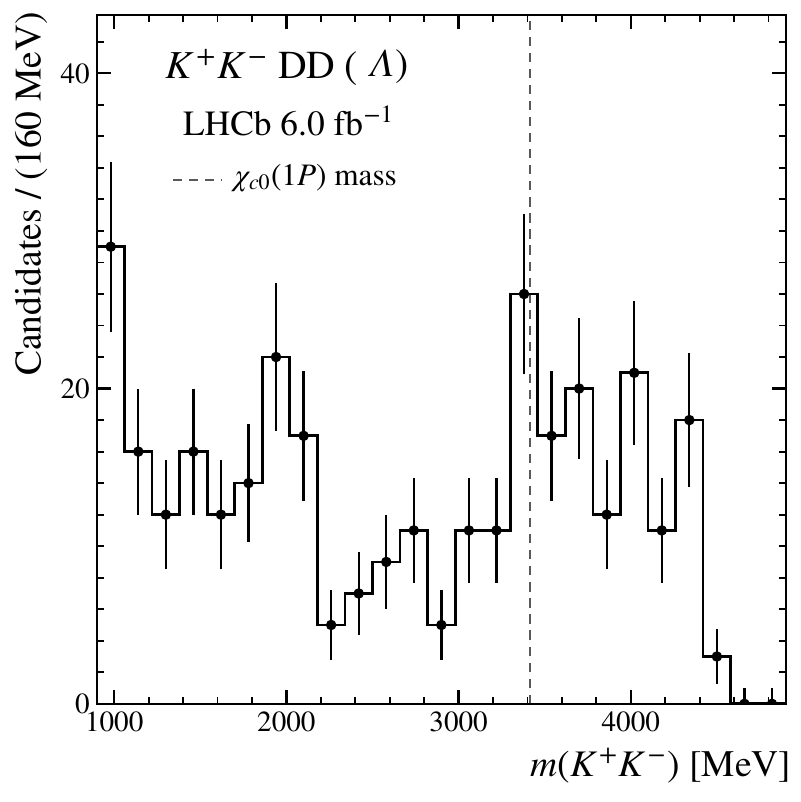}\hspace{0.02\textwidth}
    \includegraphics[width=0.42\textwidth]{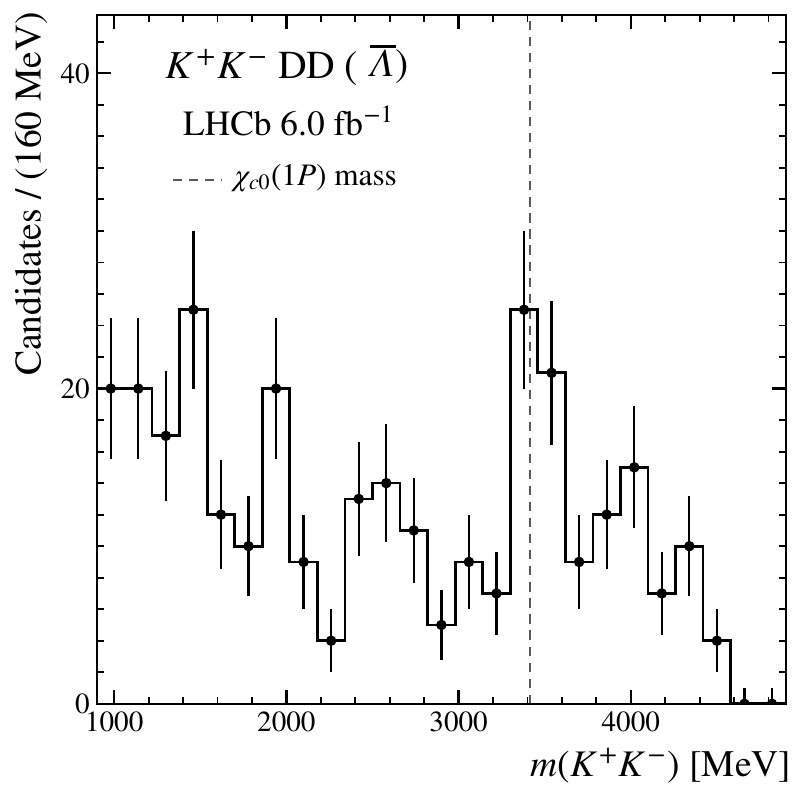}\\[0.2cm]

    % Bottom row
    \includegraphics[width=0.42\textwidth]{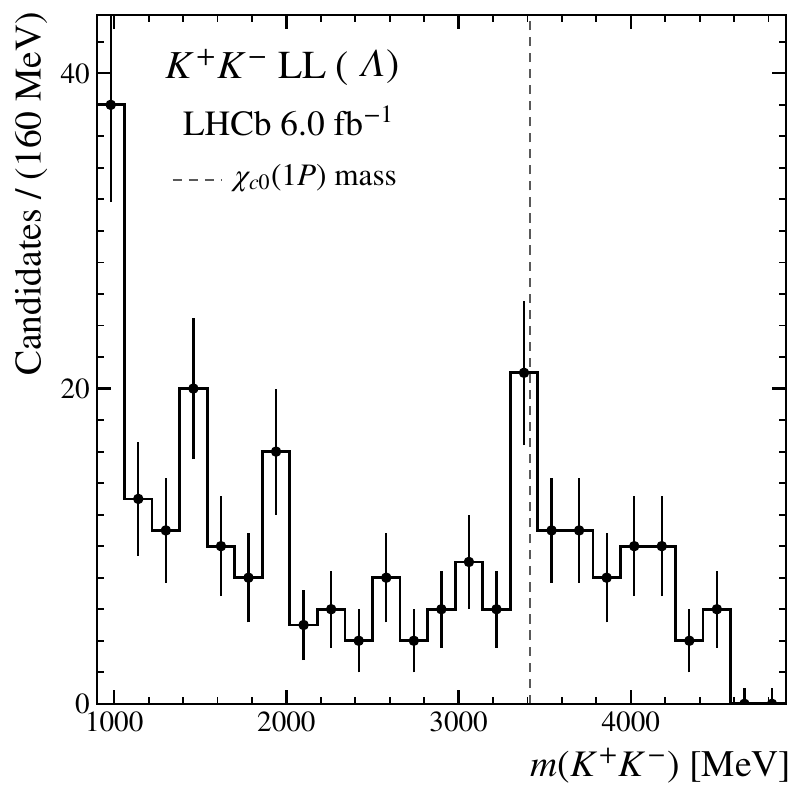}\hspace{0.02\textwidth}
    \includegraphics[width=0.42\textwidth]{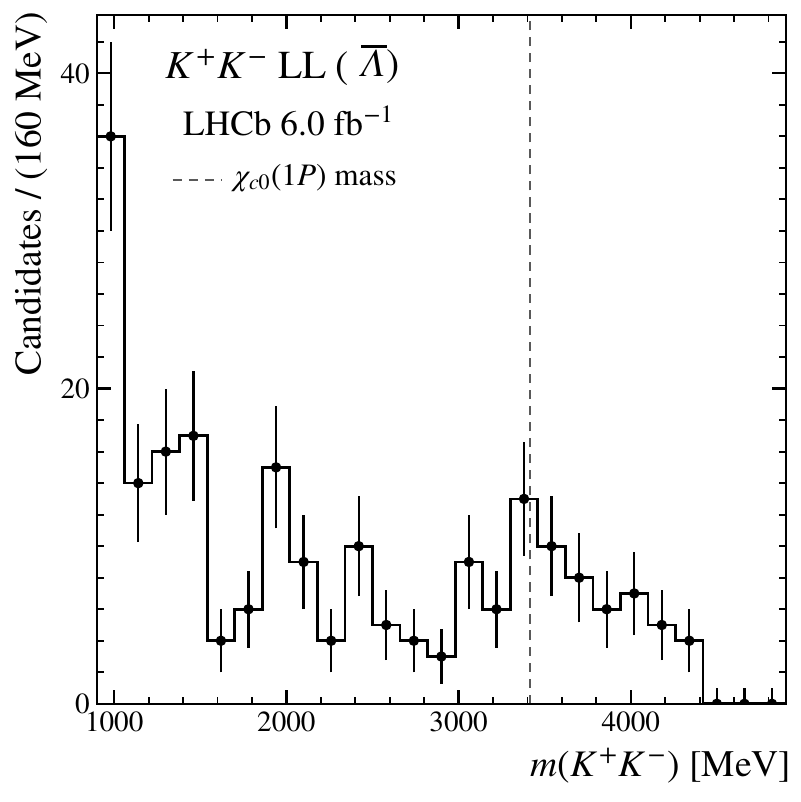}

    \caption{Distributions of the reconstructed $K^+ K^-$ invariant mass in the combined Run~2 sample of $\mathinner{\mathit{\Lambda}^0_b\!\to \mathit{\Lambda} K^+ K^-}$ candidates, selected in the $m(\mathit{\Lambda} K^+ K^-)$ signal region, defined as $m_{\mathit{\Lambda}^0_b}\pm 1.5\sigma_m$, where
$\sigma_m = 12\,\text{MeV}$.}
    \label{fig:Charmonium}
\end{figure}

Several backgrounds involving charm hadrons are suppressed with vetoes on the relevant two-body invariant-mass combinations listed in Table~\ref{tab:vetoes}. These vetoes remove contributions from decays with intermediate charm hadrons, including modes that genuinely populate the reconstructed \LbLKK final state, as well as backgrounds due to decays where final-state particles are misidentified. For the \LbLKK mode, vetoes are imposed on the invariant masses $m(\Kp\Km)$ and $m(\Lz\Kp)$ to reject contributions from intermediate \Dz, \Lc and \Xicp states. In addition, invariant masses are recomputed under alternative mass hypotheses, where one of the kaon candidates is assigned the pion mass (considering both $K^+\pi^-$ and $\pi^+K^-$ combinations), and these are used to veto residual charm backgrounds, ensuring that decays such as
$\Dz \to \Km \pip$ are rejected. The vetoes reject candidates within $\pm 30\mev$ around the known charm-hadron masses, corresponding to approximately four times the mass resolution of the relevant two-body combinations. A further requirement, $m(\Kp\Km) < 2.85\gev$, is imposed to exclude the
charmonium region. This removes the excess observed around the
$\chiczero(1P)$ mass in the normalisation mode (see Fig.~\ref{fig:Charmonium}) and suppresses possible
contamination from decays such as \LbLChiczOneP and other
\cquark\cquarkbar resonances. The same requirement is applied to the \PPbar system to remove contributions from charmonium resonances.

Less than 0.5\% of the events in both data and simulation contain more than one reconstructed \Lb candidate. To avoid correlations and ensure a uniform statistical
treatment, a single candidate per event is chosen randomly and retained. This procedure has a negligible impact on the selection
efficiency and introduces no bias in the kinematic distributions.

In addition to the combinatorial and charm backgrounds described above, additional background contributions are studied in detail using simulated samples and data from the mass sidebands. Several sources are considered, including
misidentified decays such as \LbLHH and \BdsLPH, additional decays of charmed hadrons producing the same visible final state, backgrounds involving
\KS mesons, \Lz candidates formed from random combinations of two tracks, and partially reconstructed
$b$-hadron decays. After the full selection, all studied background contributions are found to be negligible within the available statistical precision, with the exception of a residual contribution from partially reconstructed \LbSzKK decays in the normalisation channel. No other background component, apart from the smooth combinatorial contribution, is found to contribute significantly in the mass region relevant for the \LbLPPbar and \LbLKK decays.

\section{Efficiencies}
\label{sec:efficiencies}

The determination of the branching-fraction ratio requires an assessment of the relative efficiencies for the \LbLPPbar and \LbLKK decays. These efficiencies include the following effects: 
acceptance of the detector, trigger, reconstruction, and full event selection, including the multivariate selection, PID criteria and the vetoes described in Sec.~\ref{sec:vetos}.

Correction factors are applied to simulation to account for known differences with respect to data in the tracking efficiency, trigger efficiency, and PID response. Tracking and trigger corrections are derived from standard \lhcb calibration samples~\cite{LHCb-DP-2013-002}. For the PID response, per-track calibration factors are applied as described in Sec.~\ref{sec:MVA_PID}, using control samples of identified hadrons~\cite{LHCb-DP-2018-001}; these factors are parametrised as functions of $p$, $\eta$, and event multiplicity.

To account for possible nonuniformities in the three-body decay kinematics
with respect to the uniform square Dalitz phase-space generation described in Sec.~\ref{sec:DetectorAndSim},
efficiency maps are constructed in 6$\times$6 bins of the square Dalitz variables using the simulated samples with all corrections implemented. After the inclusion of the initially excluded signal region, background-subtracted Dalitz
distributions are then obtained in data using the mass as the discriminating variable (see Sec.~\ref{sec:SignalYield}) by means of the
\sPlot~\cite{Pivk:2004ty} technique, yielding per-candidate weights for the
\LbLPPbar signal. The phase-space-averaged
efficiency for each \Lz category is computed by folding these
$s$Weighted distributions with the corresponding efficiency maps. The same
procedure is applied to the normalisation mode using its much larger sample.

\begin{figure}[t]
    \centering
    % Top row
\includegraphics[width=0.48\textwidth]{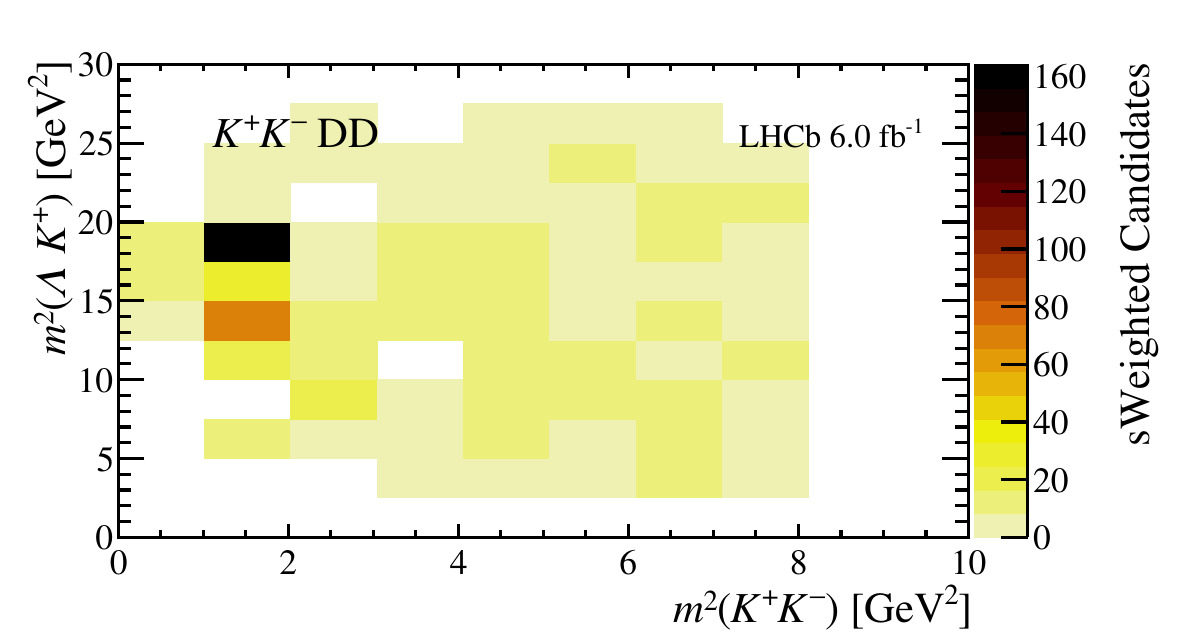}%
    %\hfill
    \includegraphics[width=0.48\textwidth]{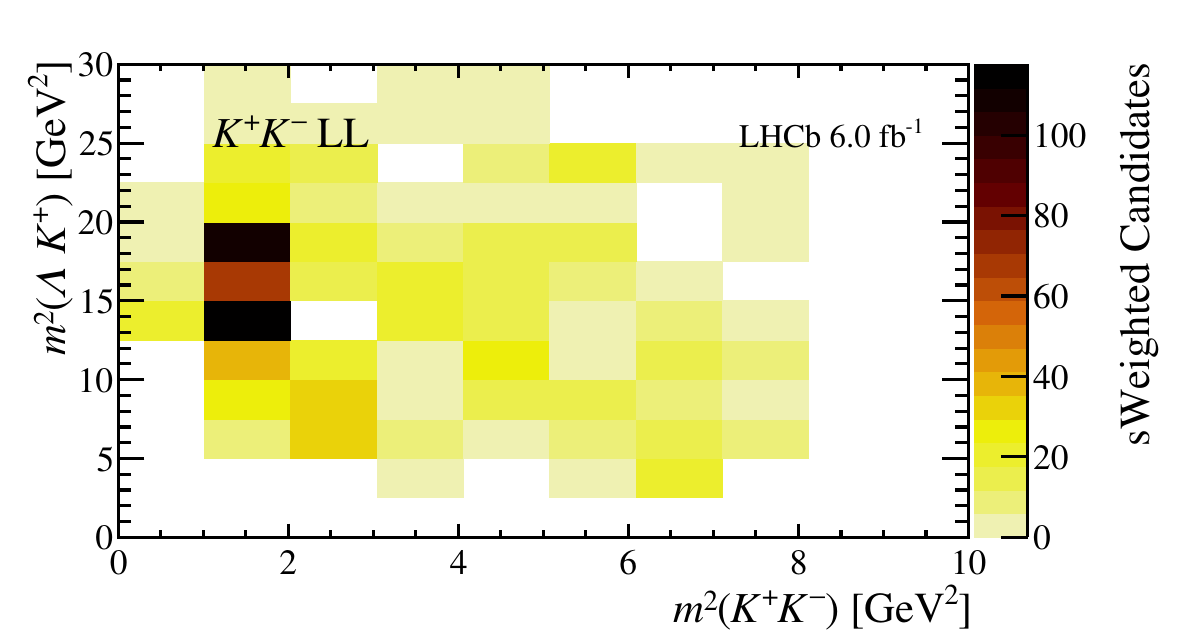}\\[0.3cm]

    % Bottom row
    \includegraphics[width=0.48\textwidth]{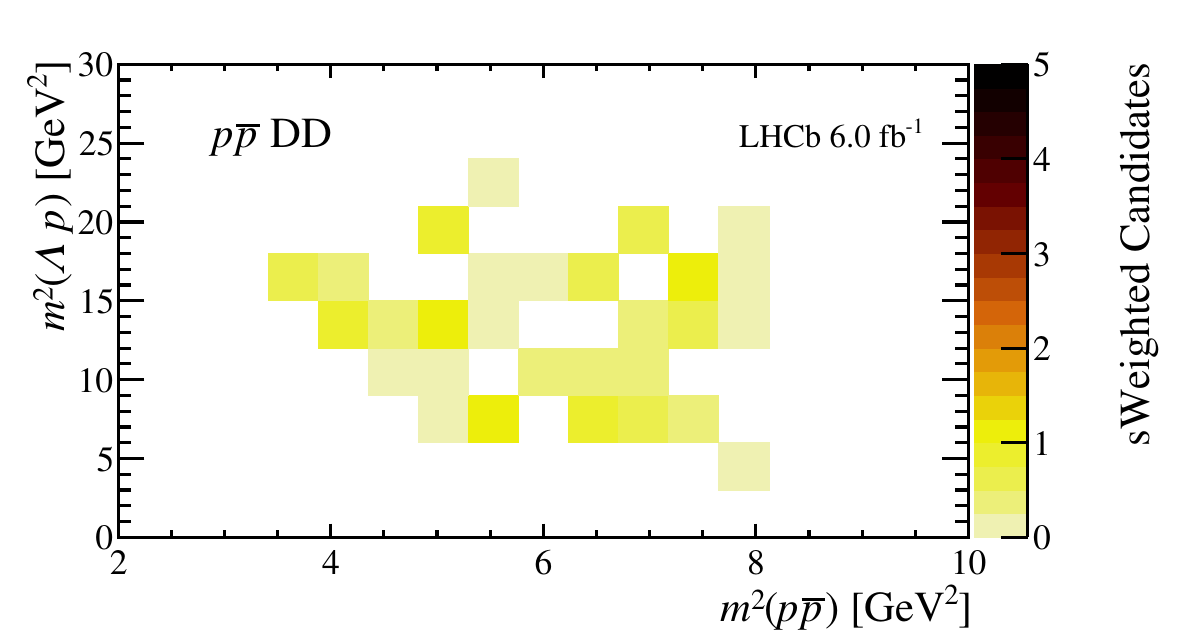}%
    %\hfill
    \includegraphics[width=0.48\textwidth]{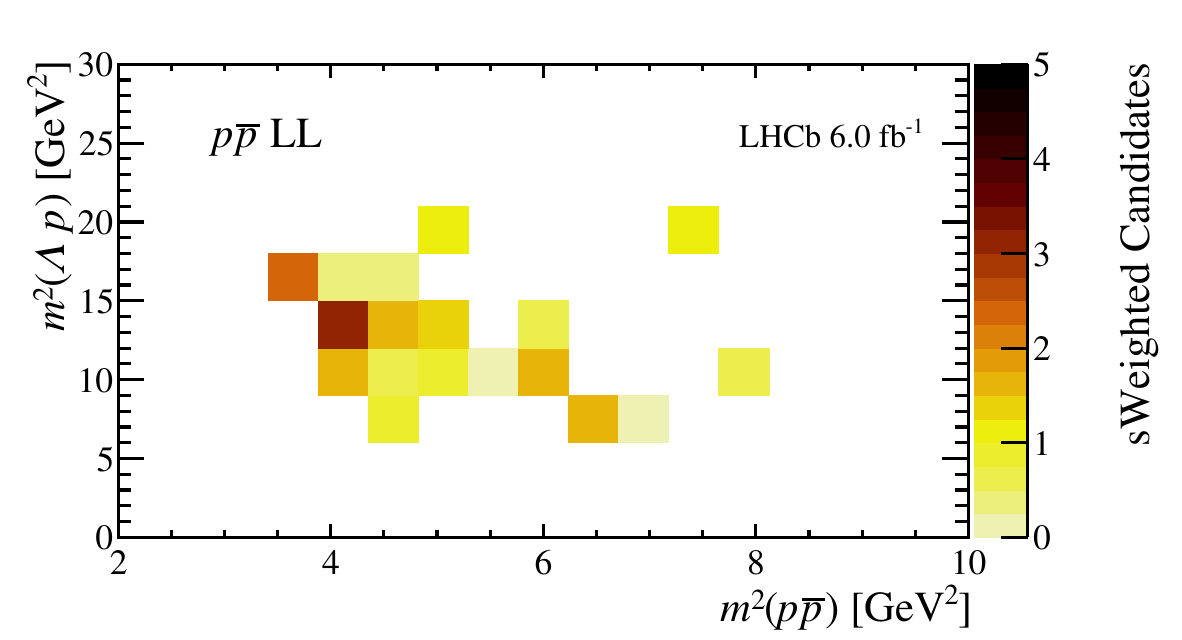}%

    \caption{Background-subtracted Dalitz plots of (top) $\mathinner{\mathit{\Lambda}^0_b\!\to \mathit{\Lambda} K^+ K^-}$ and (bottom) $\mathinner{\mathit{\Lambda}^0_b\!\to \mathit{\Lambda} p \overline{p}}$ candidates in the (left) DD and (right) LL categories in Run~2 data. The distributions are obtained after the full selection, including the requirement $m(h\bar{h})<2.85\gev$.}
    \label{fig:DalitzLbLpp}
\end{figure}

The phase-space-averaged efficiency ratios obtained from the $s$Weighted Dalitz distributions differ by about 14\% from those obtained under a uniform assumption. This difference reflects the effect of the phase-space weighting and illustrates the importance of accounting for the observed phase-space population. The uncertainty associated with the finite populations
of the efficiency-map bins is propagated through the efficiency-ratio
calculation and included in the systematic uncertainty as discussed in Sec.~\ref{sec:SystematicUncertainties}.
The $s$Weighted Dalitz distributions for the signal mode are shown in
Fig.~\ref{fig:DalitzLbLpp}. For clarity, these are presented in the standard Dalitz variables, which provide a more intuitive visualisation of the kinematic structure of the decay, while the efficiency corrections are performed using the corresponding squared Dalitz variables. The resulting phase-space-averaged efficiency
ratios for the two track categories are then used as external inputs
to the fit used to determine the signal yield, where these same quantities enter as
Gaussian-constrained nuisance parameters, as described in
Sec.~\ref{sec:SignalYield}.

\section{Signal yield determination}
\label{sec:SignalYield}
The signal yields and the branching-fraction ratio are obtained from extended
unbinned maximum-likelihood fits to the invariant-mass distributions of the
\LbLPPbar and \LbLKK candidates. The fit model described below is used for the initial fit, from which the signal $s$Weights entering the efficiency correction described in Sec.~\ref{sec:efficiencies} are determined, and for the final simultaneous fit from which the branching-fraction ratio is determined.
The fit is performed simultaneously over the LL and DD \Lz categories,
the \LbLPPbar and \LbLKK decay modes, and the data collected in the four data taking periods, providing a coherent treatment of common shape
parameters. In the final fit, the phase-space-averaged efficiency ratios entering the fit,
however, are determined as combined global quantities for each decay mode and
\Lz category, owing to the limited \LbLPPbar signal statistics.

For the normalisation mode, the fit model includes the \LbLKK signal, a partially reconstructed \LbSzKK contribution, and combinatorial background. The \Sigmaz contribution is modelled with a template
constructed from a kinematic simulation of \LbSzKK decays with \decay{\Sigmaz}{\Lz \gamma}, where the photon is assumed to be unreconstructed. The resulting $\Lambda K^+ K^-$ invariant mass distribution, including detector-resolution effects, is used as the fit template for this background component. The combinatorial background is described by an exponential function.
For the signal channel, only the \LbLPPbar and \XibzLPPbar signal components and a combinatorial background are considered.
The \XibzLPPbar component is included since it leads to the same visible final
state and would appear as a second, displaced peak in the \LPPbar
invariant-mass distribution.

Signal shapes are modelled by double-sided Crystal Ball (CB) functions~\cite{Skwarnicki:1986xj}, with the tail parameters and component fractions fixed from simulation. In particular, the \Lb peak positions are shared between the two channels, the core
width in the \LPPbar mode is taken to be proportional to that in the normalisation mode
with the proportionality factor fixed from simulation, and the \Xibz peak position is fixed relative to that of the \Lb peak using the known mass difference~\cite{PDG2024}.

Within each \Lz category, the yields in the different years are expressed
as fractions of a common total yield. The ratios of total efficiencies
between the signal and normalisation modes in the LL and DD
categories are constrained with Gaussian priors to parameter values based on simulation and corrections obtained from the data. The parameter of interest is the branching-fraction ratio between the two decay modes. The total \LbLPPbar signal yield in each track category is expressed in terms of the branching-fraction ratio,  the yield of the \LbLKK normalisation channel and the efficiency ratio. The combinatorial background in each channel and
\Lz category is described by a single exponential with a common slope
across data-taking periods. 

\begin{figure}[tb]
    \centering
    \begin{center}
        \includegraphics[width=0.445\linewidth]{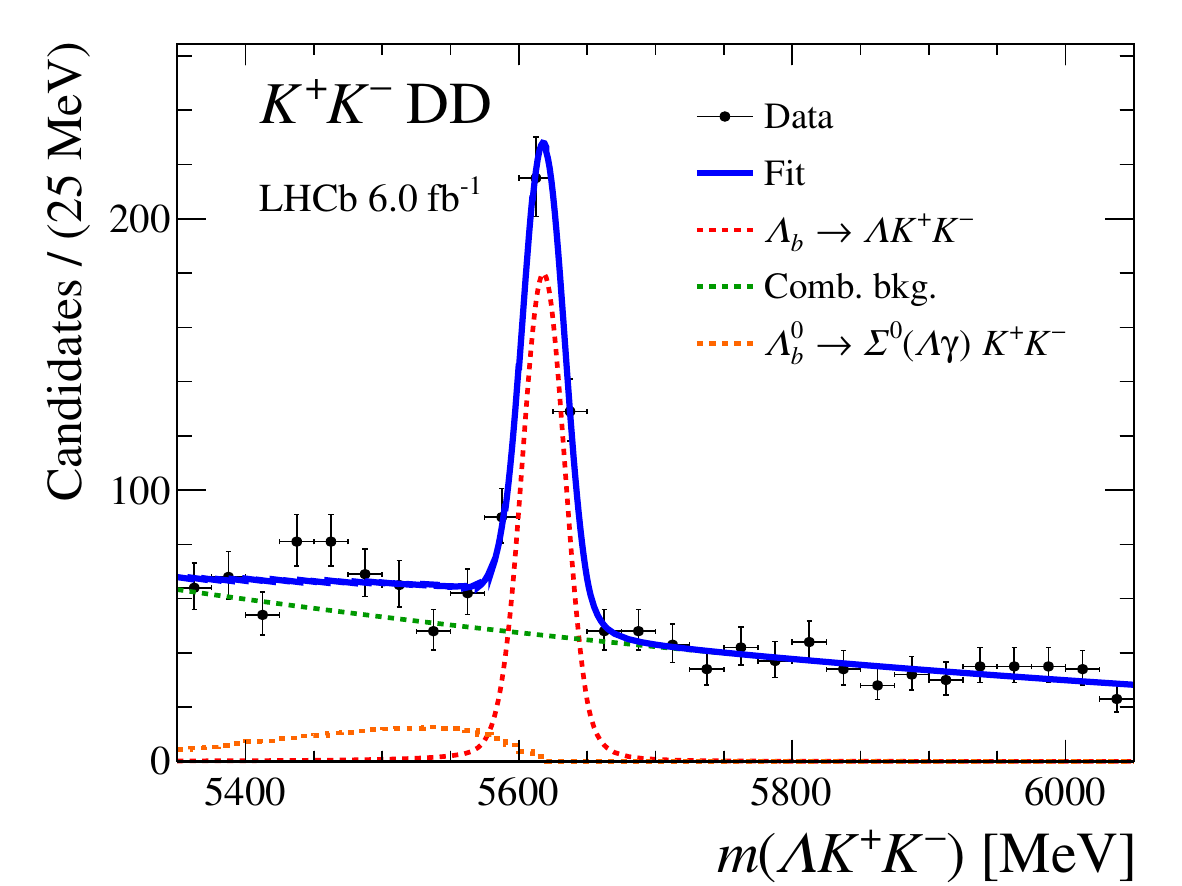}
        \includegraphics[width=0.445\linewidth]{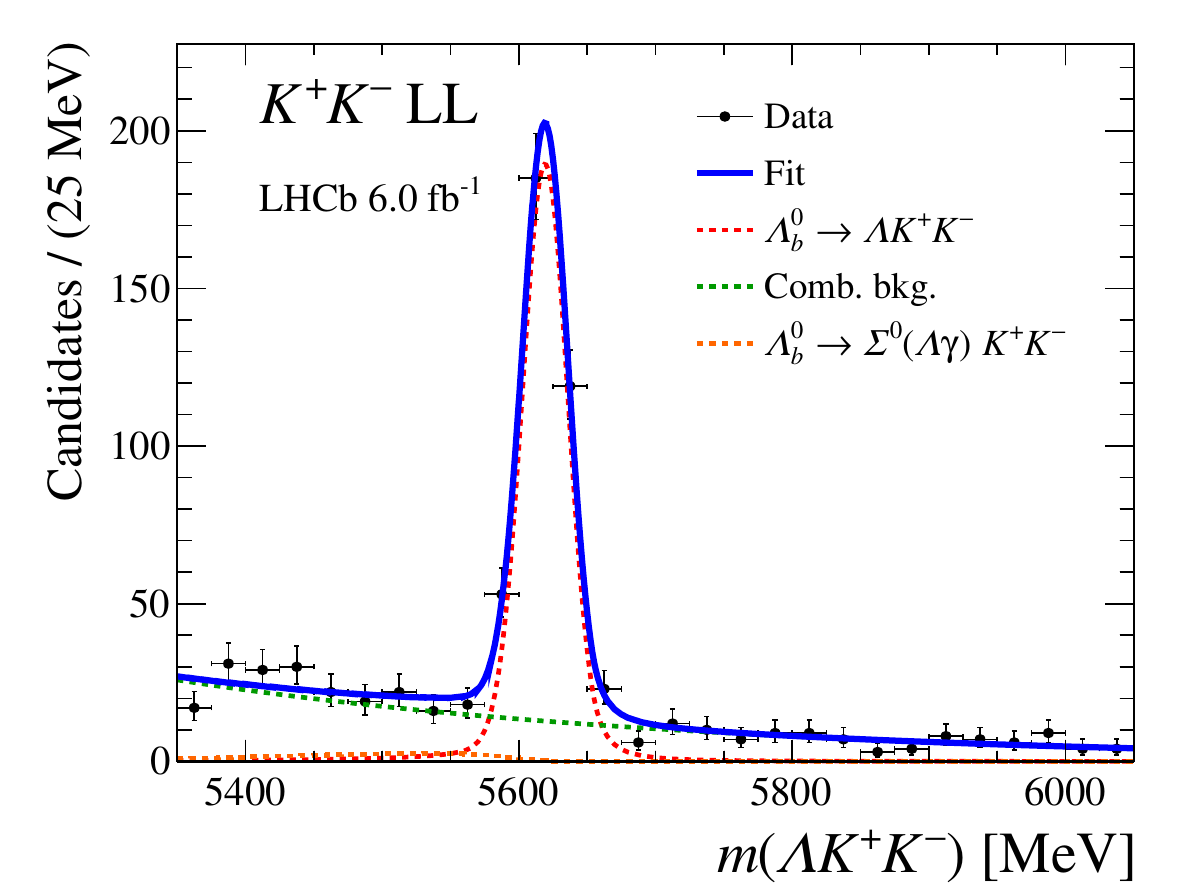}
        \includegraphics[width=0.445\linewidth]{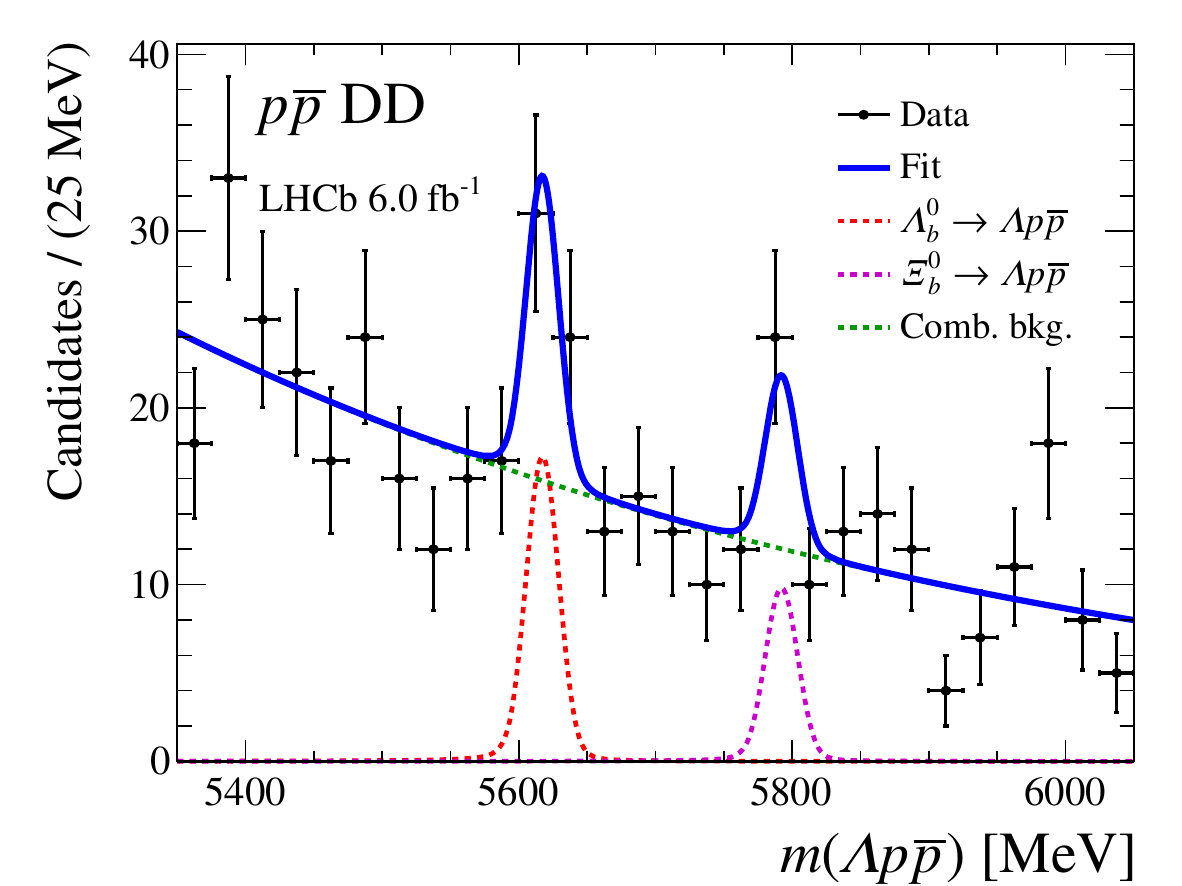}
        \includegraphics[width=0.445\linewidth]{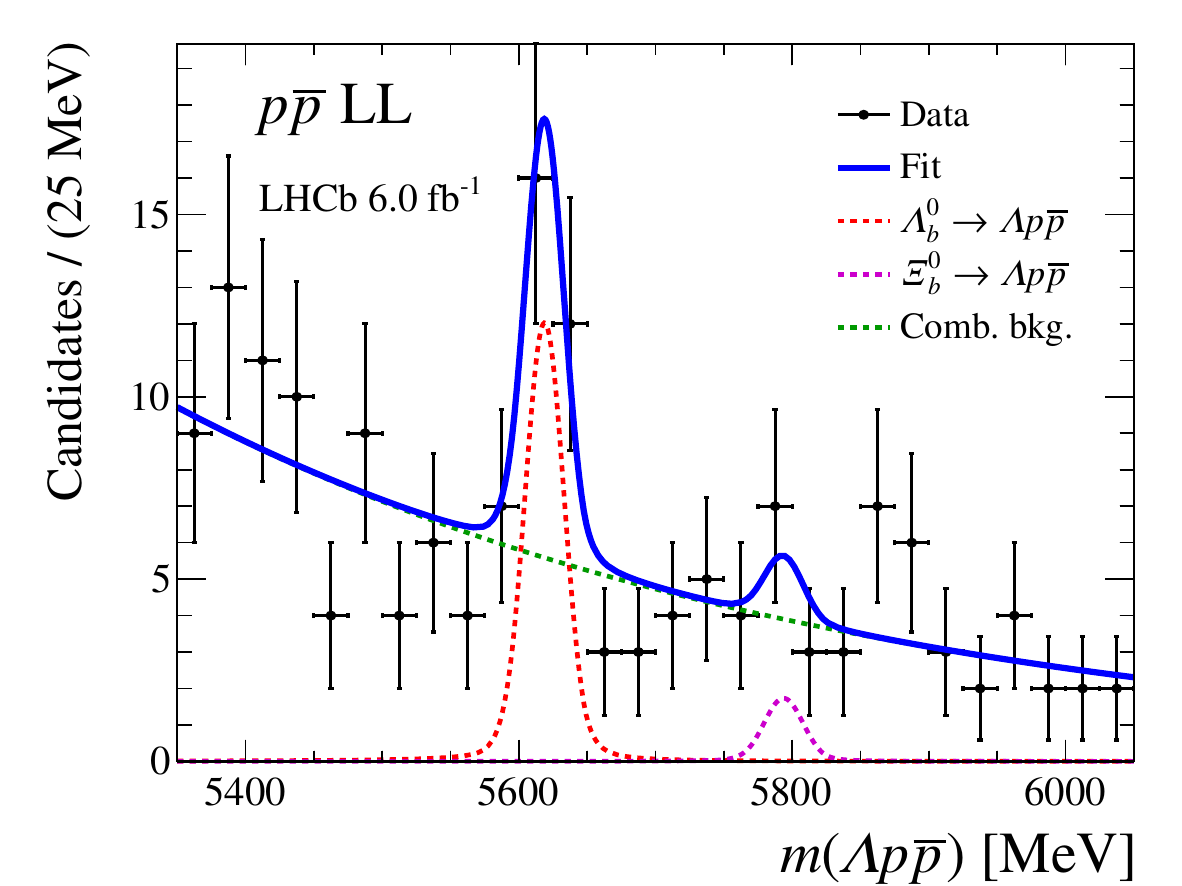}
    \end{center}
    \caption{Invariant-mass distributions of (top) $\mathinner{\mathit{\Lambda}^0_b\!\to \mathit{\Lambda} K^+ K^-}$ and (bottom) $\mathinner{\mathit{\Lambda}^0_b\!\to \mathit{\Lambda} p \overline{p}}$ 
     candidates in the (left) DD and (right) LL categories in Run~2 data with the full selection applied. The fit model is shown as a solid line.}
    \label{fig:SignalFits}
\end{figure}

This model provides a stable and unbiased description of all categories, as validated through simulation studies and fits to the control decay mode samples. The simulation-derived signal shapes and background models were validated in dedicated fits to the normalisation mode prior to the final simultaneous fit. The resulting fit projections for the \LbLKK normalisation mode and the \LbLPPbar signal mode are shown in Fig.~\ref{fig:SignalFits}.

A clear \LbLPPbar decay signal is observed for the first time. The statistical significance, computed from the likelihood-ratio test statistic $\sqrt{2\Delta\ln L}$ under the null hypothesis of zero signal contribution, is $\chi^2_{\mathrm{stat}}(0)=5.2\sigma$. The effect of systematic uncertainties is incorporated by modifying the test statistic to account for the total uncertainty. In particular, the statistical and systematic uncertainties are combined as follows
\begin{equation}
\chi^2_{\mathrm{Tot}}(0) = \frac{\chi^2_{\mathrm{stat}}(0)}{1 + \chi^2_{\mathrm{stat}}(0)\cdot \frac{\sigma^2_{\mathrm{syst}}}{(0 - S_{\mathrm{fit}})^2}} \,,
\end{equation}
where $S_{\mathrm{fit}}$ is the fitted signal yield and $\sigma_{\mathrm{syst}}$ the total systematic uncertainty. The resulting significance after including the systematic effects discussed in Sec.~\ref{sec:SystematicUncertainties} is $5.1\sigma$. 

The simultaneous fit to the data samples yields 
\( N(\LbLPPbar) = 39 \pm 10 \)
and \mbox{\( N(\LbLKK) = 640 \pm 31\)}, corresponding to fitted yields of \(20 \pm 8\) (DD) and \(19 \pm 6\) (LL) for \LbLPPbar decays.

The resulting branching-fraction ratios for each \Lz category are
\[
\begin{aligned}
R_{\Lb}^{\rm DD} &= (4.8 \pm 1.9)\times 10^{-2},\\
R_{\Lb}^{\rm LL} &= (5.4 \pm 1.7)\times 10^{-2}.
\end{aligned}
\]
The combined value
\[
R_{\Lb} = (5.1 \pm 1.3)\times 10^{-2}
\]
is obtained from a simultaneous fit to the DD and LL samples.

The quoted uncertainties on the fitted yields and branching-fraction ratios correspond to the post-fit uncertainties returned by the likelihood fit, and therefore include both the statistical contribution from the invariant-mass spectra and the effect of the Gaussian-constrained efficiency-ratio nuisance parameters.

Regarding the \XibzLPPbar decay, a small excess with a significance of $2.3 \sigma$ is observed. 

\section{Systematic uncertainties}
\label{sec:SystematicUncertainties}
Several sources of systematic uncertainty affect the measurement of the
branching-fraction ratio, as summarised in Table~\ref{tab:systematics}. They fall into three broad categories:
uncertainties related to the mass‐fit model, to the limited size of the
simulated samples, and to the residual differences between data and simulation
in the selection efficiencies.

\begin{table}[b]
    \centering
    \caption{Sources of systematic uncertainties affecting the $\mathcal{B}(\mathinner{\mathit{\Lambda}^0_b\!\to \mathit{\Lambda} p \overline{p}}) /
    \mathcal{B}(\mathinner{\mathit{\Lambda}^0_b\!\to \mathit{\Lambda} K^+ K^-})$ measurement.
    The total uncertainty is obtained by adding in quadrature all the contributions.}
    \label{tab:systematics}
    \begin{tabular}{l c}
        \hline\hline
        Source of uncertainty & Uncertainty [\%] \\
        \hline
        Fit modelling                       & 2.8 \\
        Simulated-sample statistics         & 1.0 \\
        Tracking efficiency                 & 4.3 \\
        PID efficiency                      & 0.8 \\
        Truth-matching                      & 0.2 \\
        \hline
        Total systematic uncertainty        & 5.3 \\
        \hline\hline
    \end{tabular}
\end{table}

Uncertainties associated with the signal and background mass models are
evaluated directly on data by repeating the fit under alternative modelling choices.
These include replacing the baseline double CB signal shape with a
two-sided Hypatia function~\cite{Santos:2013gra}, the exponential combinatorial-background model with a second-order Chebychev polynomial, and relaxing the assumption of common combinatorial-background slopes across data-taking periods.  The resulting variations in the fitted
branching-fraction ratio are taken as systematic uncertainties.

In addition, pseudoexperiments are generated from the baseline extended fit
results and fitted with the nominal model.
For each pseudoexperiment, samples corresponding to the different data categories are generated with event yields fluctuated according to their expected statistical uncertainties, and are subsequently fitted using the baseline procedure.
This procedure probes the behaviour of the full fit as an estimator, including possible biases arising from finite-sample
fluctuations and parameter correlations. The distribution of the fitted
branching-fraction ratios obtained from the pseudoexperiments is constructed,
and the difference between its mean value and the baseline result is taken as a
systematic uncertainty associated with the fit procedure. The total systematic
uncertainty from all fit-modelling-related sources is evaluated to be 2.8\%.

The finite size of the simulated samples used for the efficiency determination,
including the finite populations of the Dalitz efficiency-map bins entering the
phase-space-weighted efficiency calculation, is found to contribute a 1.0\%
uncertainty.  Residual data–simulation
differences in the selection efficiencies are assessed separately for
tracking, PID, and truth-matching requirements.  
Tracking efficiency uncertainties --- assigned following the \lhcb prescription of Ref.~\cite{LHCb-DP-2013-002} --- contribute 4.3\% due to hadronic interactions, 4.0\% coming from the protons and 1.5\% from the kaons. The PID
uncertainties, obtained by combining the effects of the limited calibration statistics and variations of the PID response parametrisation, contribute
0.8\%. The impact of the truth-matching requirement applied to simulated samples is
assessed by recalculating the signal and normalisation efficiency ratios after
accounting for the fraction of non-truth-matched candidates, and by verifying
that the mass distribution of such candidates is adequately described by the
signal model. The resulting effect on the measured branching-fraction ratio is
small, and a systematic uncertainty of 0.2\% is assigned.
No additional systematic uncertainty is assigned to the hardware-trigger efficiency,
since a dedicated correction is applied to account for data--simulation differences and residual effects largely cancel in the signal-to-normalisation efficiency ratio.

As the systematic sources are largely uncorrelated, their contributions are added in quadrature to obtain the total systematic uncertainty of approximately 5\%.

\section{Conclusion}
\label{sec:Conclusion}

A search for the rare charmless purely baryonic decay 
\LbLPPbar is performed using the full Run~2
\lhcb dataset.
The analysis strategy is validated using the \LbLKK decay, that also serves as normalisation mode.
After applying the full selection, including multivariate and PID
requirements and charm-hadron vetoes, the signal yield is determined from
a simultaneous fit to the \Lz categories.

The \LbLPPbar decay is observed with a
significance of $5.1 \sigma$, constituting the first observation of this mode.
The branching fraction relative to that of the topologically similar decay \LbLKK is measured excluding contributions from intermediate charmonium resonances decaying to the \PPbar and \KK final states with the requirement on the invariant mass of the hadronic system $m(h\bar{h}) < 2.85\gev$, where $h$ denotes a proton or a charged kaon. The resulting ratio of branching fractions is found to be
\[
\frac{\mathcal{B}(\LbLPPbar)}
     {\mathcal{B}(\LbLKK)}
 = (5.1 \pm 1.3_{\rm(stat)} \pm 0.3_{\rm(syst)}) \times 10^{-2}.
\]
The systematic uncertainty includes contributions from the mass-fit
modelling, efficiency corrections, PID calibration, tracking
efficiencies and limited simulation samples, as detailed in
Sec.~\ref{sec:SystematicUncertainties}.
The decay \LbLPPbar is found to be suppressed with respect to \LbLKK according to the predictions of Ref.~\cite{PBD1}.
 Since the branching fraction of the normalisation mode was measured without the invariant-mass requirement~\cite{LHCb-PAPER-2016-004}, an absolute branching fraction for the signal mode cannot be directly determined.

 A small excess with a significance of $2.3 \sigma$ is reported, consistent with contributions from \XibzLPPbar decays.
No branching fraction or upper limit is reported for this mode, since it would require a dedicated
efficiency determination including its phase-space dependence and knowledge of the
relative hadronisation fraction of \Xibz and \Lb baryons, which is beyond the
scope of this work.
 
A future analysis of the \LPPbar spectrum with the large data samples being recorded by the upgraded \lhcb detector during the LHC Run~3 will be able to study in more detail this decay mode and search for the similar mode \XibzLPPbar.
Of particular interest is the study of the two-body spectra and the likely exhibition of an enhancement at threshold~\cite{PBD1}.

% Comment this in for paper drafts; do not include this in analysis note, conference and figure reports
\section*{Acknowledgements}
%
% These Acknowledgements valid from 21/02/26
%
\noindent We express our gratitude to our colleagues in the CERN
accelerator departments for the excellent performance of the LHC. We
thank the technical and administrative staff at the LHCb
institutes.
We acknowledge support from CERN and from the national agencies:
ARC (Australia);
CAPES, CNPq, FAPERJ and FINEP (Brazil); 
MOST and NSFC (China); 
CNRS/IN2P3 and CEA (France);  % added CEA 26/02/2026
BMFTR, DFG and MPG (Germany);
INFN (Italy); 
NWO (Netherlands); 
MNiSW and NCN (Poland); 
MEC/IFA (Romania); 
%MSHE (Russia); 
MICIU and AEI (Spain);
SNSF and SER (Switzerland); 
NASU (Ukraine); 
STFC (United Kingdom); 
DOE NP and NSF (USA).
%%%%%%%%%%%%%%%%%%%%%%%%%%%%%%%%%%%%%%%%%%%%%
We acknowledge the computing resources that are provided by ARDC (Australia), 
CBPF (Brazil),
CERN, 
IHEP and LZU (China),
IN2P3 (France), 
KIT and DESY (Germany), 
INFN (Italy), 
SURF (Netherlands),
Polish WLCG (Poland),
IFIN-HH (Romania), % http://dx.doi.org/10.13039/100019931,"Institutul National de Cercetare-Dezvoltare pentru Fizica si Inginerie Nucleara 'Horia Hulubei'"
%RRCKI and Yandex LLC (Russia), 
PIC (Spain), CSCS (Switzerland), 
GridPP (United Kingdom),
and NSF (USA).  % added Feb2026
%%%%%%%%%%%%%%%%%%%%%%%%%%%%%%%%%%%%%%%%%% 
We are indebted to the communities behind the multiple open-source
software packages on which we depend.
%%%%%%%%%%%%%%%%%%%%%%%%%%%%%%%%%%%%%%%%%%
Individual groups or members have received support from
% ARC and ARDC (Australia); % moved to national 16/01/2025
RTP (Australia), % added 06/03/2026
FWO Odysseus grant G0ASD25N (Belgium), % added 20/4/2026
Key Research Program of Frontier Sciences of CAS, CAS PIFI, CAS CCEPP (China); 
%Fundamental Research Funds for the Central Universities,  and Sci.\
%\& Tech.\ Program of Guangzhou (China); Removed 24/11/25
Minciencias (Colombia);
EPLANET, Marie Sk\l{}odowska-Curie Actions, ERC and NextGenerationEU (European Union);
A*MIDEX, ANR, IPhU and Labex P2IO, and R\'{e}gion Auvergne-Rh\^{o}ne-Alpes (France);
%RFBR, RSF and Yandex LLC (Russia);
Alexander-von-Humboldt Foundation (Germany);
ICSC (Italy); 
%GVA, XuntaGal, GENCAT, Inditex, InTalent and Prog.~Atracci\'on Talento, CM (Spain);
Severo Ochoa and Mar\'ia de Maeztu Units of Excellence, GVA, XuntaGal, GENCAT, InTalent-Inditex and Prog.~Atracci\'on Talento CM (Spain);
%XuntaGal --> Xunta de Galicia 
% SRC (Sweden);  % removed 27/02/2026 - end of grant
the Leverhulme Trust, the Royal Society and UKRI (United Kingdom).

\addcontentsline{toc}{section}{References}
%\setboolean{inbibliography}{true}
\bibliographystyle{LHCb/LHCb}
\bibliography{main,LHCb/standard,LHCb/LHCb-PAPER,LHCb/LHCb-CONF,LHCb/LHCb-DP,LHCb/LHCb-TDR,LHCb/LHCb-PUB}

\newpage
% LHCb collaboration author list
% Data extracted on April 27th, 2026 at 11:35am for paper reference LHCb-PAPER-2026-004
\centerline
{\large\bf LHCb collaboration}
\begin
{flushleft}
\small
R.~Aaij$^{38}$\lhcborcid{0000-0003-0533-1952},
M.~Abdelfatah$^{69}$,
A.S.W.~Abdelmotteleb$^{57}$\lhcborcid{0000-0001-7905-0542},
C.~Abellan~Beteta$^{51}$\lhcborcid{0009-0009-0869-6798},
F.~Abudin\'en$^{59}$\lhcborcid{0000-0002-6737-3528},
T.~Ackernley$^{61}$\lhcborcid{0000-0002-5951-3498},
A.A.~Adefisoye$^{69}$\lhcborcid{0000-0003-2448-1550},
B.~Adeva$^{47}$\lhcborcid{0000-0001-9756-3712},
M.~Adinolfi$^{55}$\lhcborcid{0000-0002-1326-1264},
P.~Adlarson$^{87,42}$\lhcborcid{0000-0001-6280-3851}
% ...

{\footnotesize \it

% ...
$^{38}$Nikhef National Institute for Subatomic Physics, Amsterdam, Netherlands\\
% ...
$^{42}$National Center for Nuclear Research (NCBJ), Warsaw, Poland\\
% ...
$^{47}$Instituto Galego de F{\'\i}sica de Altas Enerx{\'\i}as (IGFAE), Universidade de Santiago de Compostela, Santiago de Compostela, Spain\\
% ...
$^{51}$Physik-Institut, Universit{\"a}t Z{\"u}rich, Z{\"u}rich, Switzerland\\
% ...
$^{55}$H.H. Wills Physics Laboratory, University of Bristol, Bristol, United Kingdom\\
% ...
$^{57}$Department of Physics, University of Warwick, Coventry, United Kingdom\\
% ...
$^{59}$School of Physics and Astronomy, University of Edinburgh, Edinburgh, United Kingdom\\
$^{60}$School of Physics and Astronomy, University of Glasgow, Glasgow, United Kingdom\\
$^{61}$Oliver Lodge Laboratory, University of Liverpool, Liverpool, United Kingdom\\
% ...
$^{69}$Syracuse University, Syracuse, NY, United States\\
% ...
$^{87}$Department of Physics and Astronomy, Uppsala University, Uppsala, Sweden, associated to $^{60}$\\
% ...
}
\end{flushleft}

\end{document}